\documentclass[10pt,a4paper,twocolumn,english,twocolumn]{IEEEtran}
\usepackage[T1]{fontenc}
\usepackage{color}
\usepackage{verbatim}
\usepackage{float}
\usepackage{units}
\usepackage{textcomp}
\usepackage{mathrsfs}
\usepackage{amsmath}
\usepackage{amsthm}
\usepackage{amssymb}
\usepackage{graphicx}
\usepackage{setspace}
\usepackage{epstopdf}

\makeatletter

\pdfpageheight\paperheight
\pdfpagewidth\paperwidth

\floatstyle{ruled}
\newfloat{algorithm}{tbp}{loa}
\providecommand{\algorithmname}{Algorithm}
\floatname{algorithm}{\protect\algorithmname}

\theoremstyle{plain}
\newtheorem{thm}{\protect\theoremname}
\theoremstyle{plain}
\newtheorem{lem}[thm]{\protect\lemmaname}

\usepackage{subfigure}
\usepackage{epstopdf}
\usepackage{cite}
\usepackage{citesort}
\usepackage{balance}

\makeatother

\usepackage{babel}
\providecommand{\lemmaname}{Lemma}
\providecommand{\theoremname}{Theorem}

\begin{document}

\title{Performance Impact of Base Station\\
Antenna Heights in Dense Cellular Networks
}
\begin{singlespace}

\author{\noindent {\normalsize{}Ming Ding, }\textit{\normalsize{}Senior Member,
IEEE}{\normalsize{}, David L$\acute{\textrm{o}}$pez P$\acute{\textrm{e}}$rez,
}\textit{\normalsize{}Senior Member, IEEE}{\normalsize{} }
}
\end{singlespace}
\maketitle
\begin{abstract}
In this paper, we present a new and significant theoretical discovery.
If the absolute height difference between base station (BS) antenna
and user equipment (UE) antenna is larger than zero, then the network
performance in terms of both the coverage probability and the area
spectral efficiency (ASE) will continuously decrease toward \emph{zero}
as the BS density increases in ultra-dense networks (UDNs). Such findings
are completely different from the conclusions in existing works, both
quantitatively and qualitatively. In particular, this performance
behavior has a tremendous impact on the deployment of the 5th-generation
(5G) UDNs. Network operators may invest large amounts of money in
deploying more network infrastructure to only obtain an even less
network capacity. Our study results reveal that one way to address
this issue is to lower the BS antenna height to the UE antenna height.
However, this requires a revolutionized approach of BS architecture
and deployment, which is explored in this paper too.
\footnote{To appear in IEEE TWC. 1536-1276 © 2015 IEEE. Personal use is permitted, but republication/redistribution requires IEEE permission. Please find the final version in IEEE from the link: http://ieeexplore.ieee.org/document/xxxxxxx/. Digital Object Identifier: 10.1109/TWC.2017.xxxxxxx}
\end{abstract}


\section{Introduction\label{sec:Introduction}}

From 1950 to 2000, the wireless network capacity has increased around
1 million fold, in which an astounding 2700\texttimes{} gain was achieved
through network densification using smaller cells~\cite{Webb_survey}.
In the first decade of 2000, network densification continued to fuel
the 3rd Generation Partnership Project (3GPP) 4th-generation (4G)
Long Term Evolution (LTE) networks, and is expected to remain as one
of the main forces to drive the 5th-generation (5G) networks onward~\cite{Tutor_smallcell}.
Indeed, the orthogonal deployment of ultra-dense (UD) small cell networks
(SCNs) within the existing macrocell network, i.e., small cells and
macrocells operating on different frequency spectrum (3GPP Small Cell
Scenario \#2a~\cite{TR36.872}), is envisaged as the workhorse for
capacity enhancement in 5G due to its large spectrum reuse and its
easy management; the latter one arising from its low interaction with
the macrocell tier, e.g., no inter-tier interference~\cite{Tutor_smallcell}.
In this paper, the focus is on the analysis of these UD SCNs with
an orthogonal deployment with the macrocells.

\begin{figure}
\noindent \centering{}\includegraphics[width=8cm]{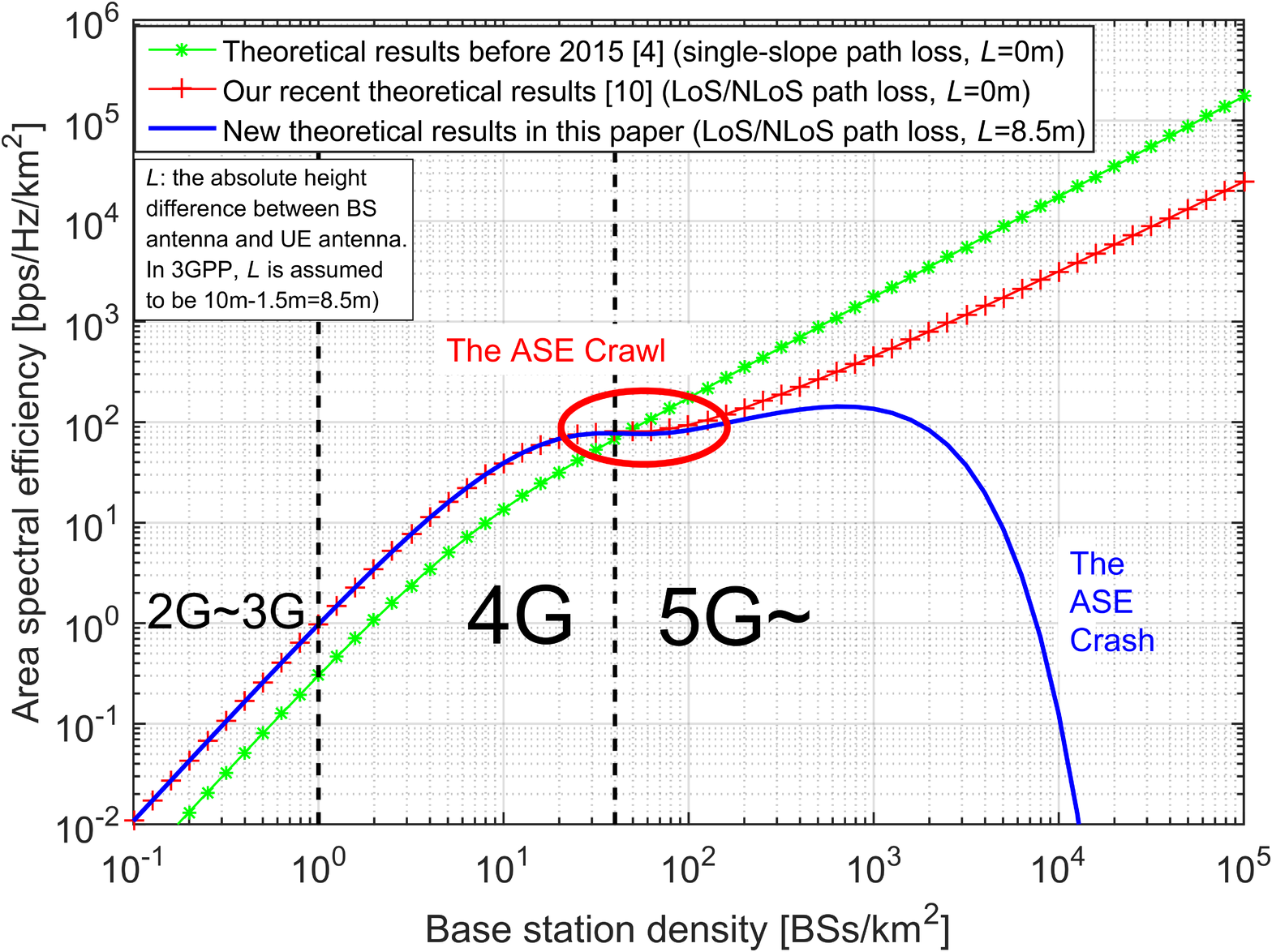}\renewcommand{\figurename}{Fig.}\caption{\label{fig:comp_ASE_2Gto5G}Theoretical comparison of the ASE performance
in $\textrm{bps/Hz/km}^{2}$. Note that all the results are obtained
using practical 3GPP channel models~\cite{TR36.828,SCM_pathloss_model},
which will be introduced later.}
\end{figure}

Before 2015, the common understanding on ultra-dense networks (UDNs)
was that the density of base stations (BSs) would not affect the per-BS
coverage probability performance in interference-limited and fully-loaded
wireless networks, and thus the area spectral efficiency (ASE) performance
in $\textrm{bps/Hz/km}^{2}$ would scale linearly with the network
densification~\cite{Jeff2011}. The implication of such conclusion
is huge: \emph{The BS density does NOT matter}, since the increase
in the interference power caused by a denser network would be exactly
compensated by the increase in the signal power due to the reduced
distance between transmitters and receivers. Hence, we do not need
to worry about the \emph{quantity} of the interferers in UDNs because
the signal power can always maintain a superior \emph{quality} over
the aggregate interference power with the network densification. Fig.~\ref{fig:comp_ASE_2Gto5G}
shows the theoretical ASE performance predicted in~\cite{Jeff2011}
across the typical BS density regions for various generations of telecommunication
systems.%
{}
 However, it is important to note that this conclusion in~\cite{Jeff2011}
was obtained with considerable simplifications on the propagation
environment, which should be placed under scrutiny when evaluating
UDNs, since they are fundamentally different from sparse ones in various
aspects~\cite{Tutor_smallcell}.

In the past year, a few noteworthy studies have been carried out to
revisit the network performance analysis for UDNs under more practical
propagation assumptions. In~\cite{related_work_Jeff}, the authors
considered a multi-slope piece-wise path loss function, while in~\cite{Related_work_Health},
the authors investigated line-of-sight (LoS) and non-line-of-sight
(NLoS) transmission as a probabilistic event for a millimeter wave
communication scenario. The most important finding in these two works
was that the per-BS coverage probability performance starts to decrease
when the BS density is sufficiently large. Fortunately, such decrease
of coverage probability will not change the monotonic increase of
the ASE as the BS density increases~\cite{related_work_Jeff,Related_work_Health}.

In our very recent work~\cite{our_GC_paper_2015_HPPP,our_work_TWC2016},
we took a step further and generalized the works in~\cite{related_work_Jeff,Related_work_Health}
by considering both piece-wise path loss functions and probabilistic
LoS/NLoS transmissions.%
{} We found that the ASE will suffer from a slow growth or even a small\emph{
decrease} on the journey from 4G to 5G when the BS density is larger
than a threshold. Fig.~\ref{fig:comp_ASE_2Gto5G} shows such theoretical
results on the ASE performance, where such threshold is around 20\,$\textrm{BSs/km}^{2}$
and the slow/negative ASE growth is highlighted by a circled area.
This circled area is referred to as \emph{the ASE Crawl} hereafter.
The intuition of \emph{the ASE Crawl} is that the aggregate interference
power increases faster than the signal power due to the transition
of a large number of interference paths from NLoS to LoS with the
network densification. The implication is profound:\emph{ The BS density
DOES matter,} since it affects the signal to interference relationship
in terms of the power ratio. Hence, in UDNs we need to pay special
attention to the \emph{quality} of the interferers, because many of
them may transit from NLoS to LoS, thus overwhelming the useful signal.
As a result, network operators should be careful when deploying dense
SCNs in order to avoid investing huge amounts of money and end up
obtaining no performance gain due to \emph{the ASE Crawl}. Fortunately,
our results in~\cite{our_GC_paper_2015_HPPP,our_work_TWC2016} also
pointed out that the ASE will again grow almost linearly as the network
further evolves to a UDN, e.g., $>10^{3}\thinspace\textrm{BSs/km}^{2}$
in Fig.~\ref{fig:comp_ASE_2Gto5G}. According to our results and
considering a 300\,MHz bandwidth, if the BS density can go as high
as $10^{4}\thinspace\textrm{BSs/km}^{2}$, the problem of \emph{the
ASE Crawl} caused by the NLoS to LoS transition of interfering paths
can be overcome, and an area throughput of $10^{3}\thinspace\textrm{Gbps/km}^{2}$
can be achieved, thus opening up an efficient way forward to 5G.

Unfortunately, the NLoS to LoS transition of interference paths is
not the only obstacle to efficient UDNs in 5G, and there are more
challenges to overcome to get there. In this paper, we present a serious
problem posed by the absolute antenna height difference between BSs
and user equipments (UEs), and evaluate its impact on UDNs by means
of a three-dimensional (3D) stochastic geometry analysis (SGA). We
made a new and significant theoretical discovery: If the absolute
antenna height difference between BSs and UEs, denoted by $L$, is
larger than zero, then the ASE performance will continuously decrease
toward \emph{zero} as the network goes ultra-dense. Fig.~\ref{fig:comp_ASE_2Gto5G}
illustrates the significance of such theoretical finding with $L=8.5$\,m\footnote{The BS antenna height and the UE antenna height are assumed to be
10\,m and 1.5\,m, respectively~\cite{TR36.814}. It is very important
to note that, compared with the existing works with $L=0$\,m~\cite{our_GC_paper_2015_HPPP,our_work_TWC2016},
our new discovery has been achieved by changing nothing but adopting
the more practical 3GPP assumption that $L=8.5$\,m~\cite{TR36.814}. }: After \emph{the ASE Crawl}, the ASE performance only increases marginally
(\textasciitilde{}1.4x) from 109.1\,$\textrm{bps/Hz/km}^{2}$ to
149.6\,$\textrm{bps/Hz/km}^{2}$ as the BS density goes from $200\thinspace\textrm{BSs/km}^{2}$
to $10^{3}\thinspace\textrm{BSs/km}^{2}$, which is then followed
by a continuous and quick fall to zero starting from around $10^{3}\thinspace\textrm{BSs/km}^{2}$.
The implication of this result is even more profound than that of
\emph{the ASE Crawl}, since following a traditional deployment with
BSs deployed at lamp posts or similar heights will dramatically reduce
the network performance in 5G. Such decline of ASE in UDNs will be
referred to as \emph{the ASE Crash} hereafter, and its fundamental
reasons will be explained in detail later in this paper.

In order to address the problem of \emph{the ASE Crash}, we propose
to change the traditional BS deployment, and lower the BS antenna
height, not just by a few meters, but straight to the UE antenna height~\cite{pre_work_ASE_crash_GC16},
so that the ASE behavior of UDNs can roll back to our previous results
in~\cite{our_work_TWC2016}, thus avoiding \emph{the ASE Crash. }However,
this brings revolutionized BS deployments and new hardware issues,
which will be discussed later.%

The rest of this paper is structured as follows. Section~\ref{sec:Related-Work}
provides a brief review on the related work. Section~\ref{sec:System-Model}
describes the system model for the 3D SGA. Section~\ref{sec:General-Results}
presents our theoretical results on the coverage probability and the
ASE performance. The numerical results are discussed in Section~\ref{sec:Simulation-and-Discussion},
with remarks shedding new light on \emph{the ASE Crash} phenomenon.
Finally, the conclusions are drawn in Section~\ref{sec:Conclusion}.

\section{Related Work\label{sec:Related-Work}}

In stochastic geometry, BS positions are typically modeled as a Homogeneous
Poisson Point Process (HPPP) on the plane, and closed-form expressions
of coverage probability can be found for some scenarios in single-tier
cellular networks~\cite{Jeff2011} and multi-tier cellular networks~\cite{Dhillon2012hetNetSG}.
The major conclusion in~\cite{Jeff2011,Dhillon2012hetNetSG} is that
neither the number of cells nor the number of cell tiers changes the
coverage probability in interference-limited fully-loaded wireless
networks. Recently, a few noteworthy studies have been carried out
to revisit the network performance analysis for dense and UDNs under
more practical propagation assumptions. As have discussed in Section~\ref{sec:Introduction},
the authors of~\cite{related_work_Jeff} and~\cite{Related_work_Health}
found that the per-BS coverage probability performance will start
to decrease when the BS density is sufficiently large%
. In our very recent work~\cite{our_GC_paper_2015_HPPP,our_work_TWC2016},
we presented a new finding that the ASE will suffer from a slow growth
or even a small\emph{ decrease} on the journey from 4G to 5G when
the BS density is larger than a threshold, i.e., \emph{the ASE Crawl}.%
{} However, none of the above works considered the antenna heights of
BSs and UEs in the theoretical analysis, which will be the focus of
this work. It is very important to note that the authors of~\cite{Renzo2015mmWave,Renzo2016intensityMatch}
recently proposed a new approach of network performance analysis based
on HPPP intensity matching. Such new approach may also be used to
investigate the BS antenna height issue, and it is interesting to
conduct a comparison study on the intensity matching approach and
our analysis in terms of the accuracy loss due to approximation and
the computational complexity, etc. In this work, we will focus on
revealing \emph{the ASE Crash} phenomenon using the traditional framework
developed in~\cite{related_work_Jeff,Related_work_Health,our_GC_paper_2015_HPPP,our_work_TWC2016}.

Another research area relating to the antenna height issue is that
of unmanned aerial vehicles (UAVs), which has attracted significant
attention as key enablers for rapid network deployment, where the
antenna heights of drone BSs and ground UEs are usually considered~\cite{Bekmezci2013FANETs,Jiang2012UAVheading,pre_work_drone_BSs_GC16,Mozaffari2015twoUAVs}.%
{} Generally speaking, the works on drone BSs put a lot of emphasis
on the 3D mobility of UAVs and try to numerically find the optimal
position/height for the drone deployment in a small area involving
just one or a few flying BSs. In contrast, our work considers a large-scale,
randomly-deployed and stationary cellular network, paying special
attention to the capacity scaling law of the whole network.%

\section{System Model\label{sec:System-Model}}

We consider a downlink (DL) cellular network with BSs deployed on
a plane according to a homogeneous Poisson point process (HPPP) $\Phi$
with a density of $\lambda$ $\textrm{BSs/km}^{2}$. Note that the
value of $\lambda$ is in the order of 10\textasciitilde{}100 for
the current 4G networks~\cite{TR36.828}. UEs are Poisson distributed
in the considered network with a density of $\rho$ $\textrm{UEs/km}^{2}$.
Note that $\rho$ is assumed to be sufficiently larger than $\lambda$
so that each BS has at least one associated UE in its coverage~\cite{related_work_Jeff,Related_work_Health,our_GC_paper_2015_HPPP,our_work_TWC2016}.
The two-dimensional (2D) distance between a BS and a UE is denoted
by $r$. Moreover, the absolute antenna height difference between
a BS and a UE is denoted by $L$. Note that the value of $L$ is in
the order of several meters. As discussed in Section~\ref{sec:Introduction},
for the current 4G networks, $L$ is around 8.5$\,$m because the
BS antenna height and the UE antenna height are assumed to be 10\,m
and 1.5\,m, respectively~\cite{TR36.814}. Hence, the 3D distance
between a BS and a UE can be expressed as
\begin{equation}
w=\sqrt{r^{2}+L^{2}}.\label{eq:actual_dis_BS2UE}
\end{equation}
Note that an alternative method is to present the 3D distance in polar
coordinates as in~\cite{Mad2014HPPP3D}.

Following~\cite{our_GC_paper_2015_HPPP,our_work_TWC2016}, we adopt
a very general path loss model, in which the path loss $\zeta\left(w\right)$
associated with distance $w$ is segmented into $N$ pieces written
as%
\begin{equation}
\zeta\left(w\right)=\begin{cases}
\zeta_{1}\left(w\right), & \textrm{when }L\leq w\leq d_{1}\\
\zeta_{2}\left(w\right), & \textrm{when }d_{1}<w\leq d_{2}\\
\vdots & \vdots\\
\zeta_{N}\left(w\right), & \textrm{when }w>d_{N-1}
\end{cases},\label{eq:prop_PL_model}
\end{equation}
where each piece $\zeta_{n}\left(w\right),n\in\left\{ 1,2,\ldots,N\right\} $
is modeled as
\begin{equation}
\zeta_{n}\left(w\right)\hspace{-0.1cm}=\hspace{-0.1cm}\begin{cases}
\hspace{-0.2cm}\begin{array}{l}
\zeta_{n}^{\textrm{L}}\left(w\right)=A_{n}^{{\rm {L}}}w^{-\alpha_{n}^{{\rm {L}}}},\\
\zeta_{n}^{\textrm{NL}}\left(w\right)=A_{n}^{{\rm {NL}}}w^{-\alpha_{n}^{{\rm {NL}}}},
\end{array} & \hspace{-0.2cm}\hspace{-0.3cm}\begin{array}{l}
\textrm{LoS:}~\textrm{Pr}_{n}^{\textrm{L}}\left(w\right)\\
\textrm{NLoS:}~1-\textrm{Pr}_{n}^{\textrm{L}}\left(w\right)
\end{array}\hspace{-0.1cm},\end{cases}\label{eq:PL_BS2UE}
\end{equation}
where $\zeta_{n}^{\textrm{L}}\left(w\right)$ and $\zeta_{n}^{\textrm{NL}}\left(w\right),n\in\left\{ 1,2,\ldots,N\right\} $
are the $n$-th piece path loss functions for the LoS transmission
and the NLoS transmission, respectively, $A_{n}^{{\rm {L}}}$ and
$A_{n}^{{\rm {NL}}}$ are the path losses at a reference distance
$w=1$ for the LoS and the NLoS cases, respectively, and $\alpha_{n}^{{\rm {L}}}$
and $\alpha_{n}^{{\rm {NL}}}$ are the path loss exponents for the
LoS and the NLoS cases, respectively. In practice, $A_{n}^{{\rm {L}}}$,
$A_{n}^{{\rm {NL}}}$, $\alpha_{n}^{{\rm {L}}}$ and $\alpha_{n}^{{\rm {NL}}}$
are constants obtainable from field tests~\cite{TR36.828,SCM_pathloss_model}.
It is very important to note that the values of the above parameters
should be chosen to ensure the continuity and the monotonically decreasing
property of the path loss function $\zeta\left(w\right)$, as the
3GPP has practiced in~\cite{TR36.828,SCM_pathloss_model}. Moreover,
$\textrm{Pr}_{n}^{\textrm{L}}\left(w\right)$ is the $n$-th piece
LoS probability function that a transmitter and a receiver separated
by a distance $w$ has a LoS path, which is assumed to be a monotonically
decreasing function with regard to $w$ in this paper.

For convenience, $\left\{ \zeta_{n}^{\textrm{L}}\left(w\right)\right\} $
and $\left\{ \zeta_{n}^{\textrm{NL}}\left(w\right)\right\} $ are
further formulated as piece-wise functions:
\begin{equation}
\zeta^{Path}\left(w\right)=\begin{cases}
\zeta_{1}^{Path}\left(w\right), & \textrm{when }L\leq w\leq d_{1}\\
\zeta_{2}^{Path}\left(w\right),\hspace{-0.3cm} & \textrm{when }d_{1}<w\leq d_{2}\\
\vdots & \vdots\\
\zeta_{N}^{Path}\left(w\right), & \textrm{when }w>d_{N-1}
\end{cases},\label{eq:general_PL_func}
\end{equation}
where the string variable $Path$ takes the value of ``L'' and ``NL''
for the LoS and the NLoS cases, respectively. Besides, $\left\{ \textrm{Pr}_{n}^{\textrm{L}}\left(w\right)\right\} $
is also formulated as a piece-wise function:
\begin{equation}
\textrm{Pr}^{\textrm{L}}\left(w\right)=\begin{cases}
\textrm{Pr}_{1}^{\textrm{L}}\left(w\right), & \textrm{when }L\leq w\leq d_{1}\\
\textrm{Pr}_{2}^{\textrm{L}}\left(w\right),\hspace{-0.3cm} & \textrm{when }d_{1}<w\leq d_{2}\\
\vdots & \vdots\\
\textrm{Pr}_{N}^{\textrm{L}}\left(w\right), & \textrm{when }w>d_{N-1}
\end{cases}.\label{eq:general_LoS_Pr}
\end{equation}

The generality and the practicality of the adopted path loss model
(\ref{eq:prop_PL_model}) have been well established in~\cite{our_work_TWC2016}.%
{} In particular, note that (\ref{eq:prop_PL_model}) can be easily
degenerated to a simplistic single-slope path loss model as~\cite{Jeff2011}
\begin{equation}
\zeta\left(w\right)=Aw^{-\alpha},\label{eq:single_slope_PL_BS2UE}
\end{equation}
which can be obtained from (\ref{eq:PL_BS2UE}) with $N=1$, $A_{1}^{{\rm {L}}}=A_{1}^{{\rm {NL}}}=A$,
$\alpha_{1}^{{\rm {L}}}=\alpha_{1}^{{\rm {NL}}}=\alpha$, and $\textrm{Pr}^{\textrm{L}}\left(w\right)=1$.
Consequently, our analytical results to be presented in later sections
can be easily applied to such simplistic single-slope path loss model
(\ref{eq:single_slope_PL_BS2UE}) with the above parameter substitution.
It should be noted that the adopted path loss model (\ref{eq:prop_PL_model})
is fundamentally different from the simplistic single-slope path loss
model (\ref{eq:single_slope_PL_BS2UE}). In more detail, for (\ref{eq:prop_PL_model}),
we assume a practical user association strategy (UAS), in which each
UE is connected to the BS with the smallest path loss (i.e., with
the largest $\zeta\left(w\right)$)~\cite{Related_work_Health,our_work_TWC2016}.
Such assumption reflects the fact that in reality a UE might connect
to a BS that is not the nearest one but with a strong LoS path, rather
than to the nearest BS with a weak NLoS path. However, for (\ref{eq:single_slope_PL_BS2UE}),
each UE is always associated with the BS at the closest proximity~\cite{Jeff2011}.%

Moreover, we assume that each BS/UE is equipped with an isotropic
antenna, and that the multi-path fading between a BS and a UE is modeled
as independently identical distributed (i.i.d.) Rayleigh fading\footnote{It is important to note that in practical 4G/5G networks~\cite{Tutor_smallcell},
multi-path fading is usually \emph{not} considered in user association
due to its fast time-varying nature, while it should be considered
in the SINR analysis. Our system model captures such key feature of
cellular networks. However, the theoretical analysis becomes challenging,
as will be shown in the next section.}~\cite{related_work_Jeff,Related_work_Health,our_GC_paper_2015_HPPP,our_work_TWC2016}.
In order to investigate more realistic channel environments, we further
perform the following studies:
\begin{itemize}
\item A practical antenna pattern and downtilt will be considered in Subsection~\ref{subsec:downtilt_impact_on_ASE}
with performance evaluated by simulations. Note that in practice each
BS antenna has a 3D beam pattern and such beam will be electrically
tilted downward to improve the signal power as well as reduce the
inter-cell interference~\cite{Li2015antenna,Fischer2002tiltAngle,Sarkar2003propagModel}.
Intuitively speaking, the downtilt angle should increase with the
network densification since each BS's effective coverage area shrinks.
For example, the downtilt angle is around 10$\,$degrees for macrocell
BSs in~\cite{TR36.814} and it is significantly larger for small
cell BSs~\cite{Tutor_smallcell}.
\item A practical Rician fading will be analyzed in Subsection~\ref{subsec:The-Analytical-Results-Rician}
and evaluated in Subsection~\ref{subsec:Rician_impact_on_ASE} to
show its minor impact on our conclusions.
\item Two practical 3GPP path loss models will be introduced as special
examples of (\ref{eq:prop_PL_model}) in Subsection~\ref{subsec:The-3GPP-Special-Cases}.
These two path loss models as well as the single-slope path loss model
(\ref{eq:single_slope_PL_BS2UE}) will be evaluated in Section~\ref{sec:Simulation-and-Discussion}
to show the generality of our results.
\end{itemize}

\section{Main Results\label{sec:General-Results}}

Using a 3D SGA based on the HPPP theory, we study the performance
of the SCN by considering the performance of a typical UE located
at the origin $o$.

\subsection{The Coverage Probability\label{subsec:The-Coverage-Probability-Performance}}

We first investigate the coverage probability $p^{{\rm {cov}}}\left(\lambda,\gamma\right)$
that the typical UE's signal-to-interference-plus-noise ratio (SINR)
is above a per-designated threshold $\gamma$:
\begin{equation}
p^{{\rm {cov}}}\left(\lambda,\gamma\right)=\textrm{Pr}\left[\mathrm{SINR}>\gamma\right],\label{eq:Coverage_Prob_def}
\end{equation}
where the SINR is calculated as
\begin{equation}
\mathrm{SINR}=\frac{P\zeta\left(w\right)h}{I_{{\rm {agg}}}+P_{{\rm {N}}}},\label{eq:SINR}
\end{equation}
where $h$ is the channel gain, modeled as an exponential random variable
(RV) with the mean of one (due to our consideration of Rayleigh fading
in this paper), $P$ is the transmission power of each BS, $P_{{\rm {N}}}$
is the additive white Gaussian noise (AWGN) power at each UE, and
$I_{{\rm {agg}}}$ is the aggregate interference given by
\begin{equation}
I_{{\rm {agg}}}=\sum_{i:\,b_{i}\in\Phi\setminus b_{o}}P\beta_{i}g_{i},\label{eq:cumulative_interference}
\end{equation}
where $b_{o}$ is the BS serving the typical UE at distance $w$ from
it, $b_{i}$ is the $i$-th interfering BS, and $\beta_{i}$ and $g_{i}$
are the path loss and the multi-path fading channel gain of $b_{i}$,
respectively.

Based on the proposed path loss model in (\ref{eq:prop_PL_model})
and the presented UAS, we present our main result on the coverage
probability $p^{{\rm {cov}}}\left(\lambda,\gamma\right)$ in Theorem~\ref{thm:p_cov_UAS1}.
\begin{algorithm*}
\begin{thm}
\label{thm:p_cov_UAS1}Considering the proposed path loss model in
(\ref{eq:prop_PL_model}) and the proposed UAS, the coverage probability
$p^{{\rm {cov}}}\left(\lambda,\gamma\right)$ can be derived as
\begin{equation}
p^{{\rm {cov}}}\left(\lambda,\gamma\right)=\sum_{n=1}^{N}\left(T_{n}^{{\rm {L}}}+T_{n}^{{\rm {NL}}}\right),\label{eq:Theorem_1_p_cov}
\end{equation}
where $T_{n}^{{\rm {L}}}\hspace{-0.1cm}=\hspace{-0.1cm}\int_{\sqrt{d_{n-1}^{2}-L^{2}}}^{\sqrt{d_{n}^{2}-L^{2}}}\hspace{-0.1cm}{\rm {Pr}}\hspace{-0.1cm}\left[\frac{P\zeta_{n}^{{\rm {L}}}\left(\sqrt{r^{2}+L^{2}}\right)h}{I_{{\rm {agg}}}+P_{{\rm {N}}}}\hspace{-0.1cm}>\hspace{-0.1cm}\gamma\right]\hspace{-0.1cm}f_{R,n}^{{\rm {L}}}\left(r\right)dr$,
$T_{n}^{{\rm {NL}}}\hspace{-0.1cm}=\hspace{-0.1cm}\int_{\sqrt{d_{n-1}^{2}-L^{2}}}^{\sqrt{d_{n}^{2}-L^{2}}}\hspace{-0.1cm}{\rm {Pr}}\hspace{-0.1cm}\left[\frac{P\zeta_{n}^{{\rm {NL}}}\left(\sqrt{r^{2}+L^{2}}\right)h}{I_{{\rm {agg}}}+P_{{\rm {N}}}}\hspace{-0.1cm}>\hspace{-0.1cm}\gamma\right]\hspace{-0.1cm}f_{R,n}^{{\rm {NL}}}\left(r\right)dr$,
and $d_{0}$ and $d_{N}$ are defined as $L$ and $+\infty$, respectively.
Moreover, $f_{R,n}^{{\rm {L}}}\left(r\right)$ and $f_{R,n}^{{\rm {NL}}}\left(r\right)$
$\left(\sqrt{d_{n-1}^{2}-L^{2}}<r\leq\sqrt{d_{n}^{2}-L^{2}}\right)$
are given by
\begin{eqnarray}
f_{R,n}^{{\rm {L}}}\left(r\right) & = & \exp\left(\hspace{-0.1cm}-\hspace{-0.1cm}\int_{0}^{r_{1}}\left(1-{\rm {Pr}}^{{\rm {L}}}\left(\sqrt{u^{2}+L^{2}}\right)\right)2\pi u\lambda du\right)\nonumber \\
 &  & \times\exp\left(\hspace{-0.1cm}-\hspace{-0.1cm}\int_{0}^{r}{\rm {Pr}}^{{\rm {L}}}\left(\sqrt{u^{2}+L^{2}}\right)2\pi u\lambda du\right){\rm {Pr}}_{n}^{{\rm {L}}}\left(\sqrt{r^{2}+L^{2}}\right)2\pi r\lambda,\hspace{-0.1cm}\hspace{-0.1cm}\hspace{-0.1cm}\hspace{-0.1cm}\label{eq:geom_dis_PDF_UAS1_LoS_thm}
\end{eqnarray}
and
\begin{eqnarray}
\hspace{-0.1cm}\hspace{-0.1cm}\hspace{-0.1cm}\hspace{-0.1cm}\hspace{-0.1cm}\hspace{-0.1cm}f_{R,n}^{{\rm {NL}}}\left(r\right) & = & \exp\left(\hspace{-0.1cm}-\hspace{-0.1cm}\int_{0}^{r_{2}}{\rm {Pr}}^{{\rm {L}}}\left(\sqrt{u^{2}+L^{2}}\right)2\pi u\lambda du\right)\nonumber \\
 &  & \times\exp\left(\hspace{-0.1cm}-\hspace{-0.1cm}\int_{0}^{r}\left(1-{\rm {Pr}}^{{\rm {L}}}\left(\sqrt{u^{2}+L^{2}}\right)\right)2\pi u\lambda du\right)\left(1-{\rm {Pr}}_{n}^{{\rm {L}}}\left(\sqrt{r^{2}+L^{2}}\right)\right)2\pi r\lambda,\hspace{-0.1cm}\hspace{-0.1cm}\hspace{-0.1cm}\hspace{-0.1cm}\label{eq:geom_dis_PDF_UAS1_NLoS_thm}
\end{eqnarray}
where $r_{1}$ and $r_{2}$ are computed implicitly by the following
equations,
\begin{equation}
r_{1}=\underset{r_{1}}{\arg}\left\{ \zeta^{{\rm {NL}}}\left(\sqrt{r_{1}^{2}+L^{2}}\right)=\zeta_{n}^{{\rm {L}}}\left(\sqrt{r^{2}+L^{2}}\right)\right\} ,\label{eq:def_r_1}
\end{equation}
and
\begin{equation}
r_{2}=\underset{r_{2}}{\arg}\left\{ \zeta^{{\rm {L}}}\left(\sqrt{r_{2}^{2}+L^{2}}\right)=\zeta_{n}^{{\rm {NL}}}\left(\sqrt{r^{2}+L^{2}}\right)\right\} .\label{eq:def_r_2}
\end{equation}
\end{thm}
\begin{IEEEproof}
See Appendix~A.
\end{IEEEproof}
\noindent \rule[0.5ex]{2\columnwidth}{1pt}
\begin{lem}
\label{lem:Pcov_LoS_and_NLoS_Rayleigh}In Theorem~\ref{thm:p_cov_UAS1},
${\rm {Pr}}\left[\frac{P\zeta_{n}^{{\rm {L}}}\left(\sqrt{r^{2}+L^{2}}\right)h}{I_{{\rm {agg}}}+P_{{\rm {N}}}}>\gamma\right]$
is given by
\begin{equation}
{\rm {Pr}}\left[\frac{P\zeta_{n}^{{\rm {L}}}\left(\sqrt{r^{2}+L^{2}}\right)h}{I_{{\rm {agg}}}+P_{{\rm {N}}}}>\gamma\right]=\exp\left(-\frac{\gamma P_{{\rm {N}}}}{P\zeta_{n}^{{\rm {L}}}\left(\sqrt{r^{2}+L^{2}}\right)}\right)\mathscr{L}_{I_{{\rm {agg}}}}^{{\rm {L}}}\left(s\right),\label{eq:Pr_SINR_req_UAS1_LoS_thm}
\end{equation}
where $s=\frac{\gamma}{P\zeta_{n}^{{\rm {L}}}\left(\sqrt{r^{2}+L^{2}}\right)}$
and
\begin{equation}
\mathscr{L}_{I_{{\rm {agg}}}}^{{\rm {L}}}\hspace{-0.1cm}\left(s\right)=\exp\hspace{-0.1cm}\left(\hspace{-0.1cm}-2\pi\lambda\hspace{-0.1cm}\int_{r}^{+\infty}\hspace{-0.1cm}\hspace{-0.1cm}\hspace{-0.1cm}\hspace{-0.1cm}\frac{{\rm {Pr}}^{{\rm {L}}}\left(\sqrt{u^{2}+L^{2}}\right)u}{1\hspace{-0.1cm}+\hspace{-0.1cm}\left(sP\zeta^{{\rm {L}}}\left(\sqrt{u^{2}+L^{2}}\right)\right)^{-1}}du\hspace{-0.1cm}\right)\exp\hspace{-0.1cm}\left(\hspace{-0.1cm}-2\pi\lambda\hspace{-0.1cm}\int_{r_{1}}^{+\infty}\hspace{-0.1cm}\hspace{-0.1cm}\hspace{-0.1cm}\hspace{-0.1cm}\frac{\left[1\hspace{-0.1cm}-\hspace{-0.1cm}{\rm {Pr}}^{{\rm {L}}}\left(\sqrt{u^{2}+L^{2}}\right)\right]u}{1\hspace{-0.1cm}+\hspace{-0.1cm}\left(sP\zeta^{{\rm {NL}}}\left(\sqrt{u^{2}+L^{2}}\right)\right)^{-1}}du\hspace{-0.1cm}\right)\hspace{-0.1cm}.\hspace{-0.1cm}\label{eq:laplace_term_LoS_UAS1_general_seg_thm}
\end{equation}

Moreover, in Theorem~\ref{thm:p_cov_UAS1}, ${\rm {Pr}}\left[\frac{P\zeta_{n}^{{\rm {NL}}}\left(\sqrt{r^{2}+L^{2}}\right)h}{I_{{\rm {agg}}}+P_{{\rm {N}}}}>\gamma\right]$
is given by
\begin{equation}
{\rm {Pr}}\left[\frac{P\zeta_{n}^{{\rm {NL}}}\left(\sqrt{r^{2}+L^{2}}\right)h}{I_{{\rm {agg}}}+P_{{\rm {N}}}}>\gamma\right]=\exp\left(-\frac{\gamma P_{{\rm {N}}}}{P\zeta_{n}^{{\rm {NL}}}\left(\sqrt{r^{2}+L^{2}}\right)}\right)\mathscr{L}_{I_{{\rm {agg}}}}^{{\rm {NL}}}\left(s\right),\label{eq:Pr_SINR_req_UAS1_NLoS_thm}
\end{equation}
where $s=\frac{\gamma}{P\zeta_{n}^{{\rm {NL}}}\left(\sqrt{r^{2}+L^{2}}\right)}$
and
\begin{equation}
\mathscr{L}_{I_{{\rm {agg}}}}^{{\rm {NL}}}\hspace{-0.1cm}\left(s\right)=\exp\hspace{-0.1cm}\left(\hspace{-0.1cm}-2\pi\lambda\hspace{-0.1cm}\int_{r_{2}}^{+\infty}\hspace{-0.1cm}\hspace{-0.1cm}\hspace{-0.1cm}\hspace{-0.1cm}\frac{{\rm {Pr}}^{{\rm {L}}}\left(\sqrt{u^{2}+L^{2}}\right)u}{1\hspace{-0.1cm}+\hspace{-0.1cm}\left(sP\zeta^{{\rm {L}}}\left(\sqrt{u^{2}+L^{2}}\right)\right)^{-1}}du\hspace{-0.1cm}\right)\exp\hspace{-0.1cm}\left(\hspace{-0.1cm}-2\pi\lambda\hspace{-0.1cm}\int_{r}^{+\infty}\hspace{-0.1cm}\hspace{-0.1cm}\hspace{-0.1cm}\hspace{-0.1cm}\frac{\left[1\hspace{-0.1cm}-\hspace{-0.1cm}{\rm {Pr}}^{{\rm {L}}}\left(\sqrt{u^{2}+L^{2}}\right)\right]u}{1\hspace{-0.1cm}+\hspace{-0.1cm}\left(sP\zeta^{{\rm {NL}}}\left(\sqrt{u^{2}+L^{2}}\right)\right)^{-1}}du\hspace{-0.1cm}\right)\hspace{-0.1cm}.\hspace{-0.1cm}\label{eq:laplace_term_NLoS_UAS1_general_seg_thm}
\end{equation}
\end{lem}
\begin{IEEEproof}
See Appendix~B.
\end{IEEEproof}
\end{algorithm*}

For clarity, ${\rm {Pr}}\left[\frac{P\zeta_{n}^{{\rm {L}}}\left(\sqrt{r^{2}+L^{2}}\right)h}{I_{{\rm {agg}}}+P_{{\rm {N}}}}>\gamma\right]$
and ${\rm {Pr}}\left[\frac{P\zeta_{n}^{{\rm {NL}}}\left(\sqrt{r^{2}+L^{2}}\right)h}{I_{{\rm {agg}}}+P_{{\rm {N}}}}>\gamma\right]$
in Theorem~\ref{thm:p_cov_UAS1} are presented in Lemma~\ref{lem:Pcov_LoS_and_NLoS_Rayleigh}.
In Theorem~\ref{thm:p_cov_UAS1}, $\mathscr{L}_{I_{{\rm {agg}}}}^{{\rm {L}}}\left(s\right)$
in (\ref{eq:Pr_SINR_req_UAS1_LoS_thm}) and $\mathscr{L}_{I_{{\rm {agg}}}}^{{\rm {NL}}}\left(s\right)$
in (\ref{eq:Pr_SINR_req_UAS1_NLoS_thm}) are the Laplace transform
of $I_{{\rm {agg}}}$ evaluated at $s$ for LoS signal transmission
and that for NLoS transmission, respectively. Regarding the computational
process to obtain $p^{{\rm {cov}}}\left(\lambda,\gamma\right)$ presented
in Theorem~\ref{thm:p_cov_UAS1}, three folds of integrals are respectively
required for the calculation of $\left\{ f_{R,n}^{Path}\left(r\right)\right\} $,
$\left\{ \mathscr{L}_{I_{{\rm {agg}}}}^{Path}\left(\frac{\gamma}{P\zeta_{n}^{Path}\left(\sqrt{r^{2}+L^{2}}\right)}\right)\right\} $,
and $\left\{ T_{n}^{Path}\right\} $, where the string variable $Path$
takes the value of ``L'' and ``NL'' for the LoS case and the NLoS
case, respectively.

\subsection{The Area Spectral Efficiency\label{subsec:The-ASE-Performance}}

According to~\cite{our_GC_paper_2015_HPPP,our_work_TWC2016}, we
also investigate the ASE in $\textrm{bps/Hz/km}^{2}$ for a given
$\lambda$, which can be computed as
\begin{equation}
A^{{\rm {ASE}}}\left(\lambda,\gamma_{0}\right)=\lambda\int_{\gamma_{0}}^{+\infty}\log_{2}\left(1+\gamma\right)f_{\mathit{\Gamma}}\left(\lambda,\gamma\right)d\gamma,\label{eq:ASE_def}
\end{equation}
where $\gamma_{0}$ is the minimum working SINR for the considered
SCN, and $f_{\mathit{\Gamma}}\left(\lambda,\gamma\right)$ is the
probability density function (PDF) of the SINR at the typical UE for
a particular value of $\lambda$. The ASE defined in this paper is
different from that in~\cite{related_work_Jeff}, where a constant
rate based on $\gamma_{0}$ is assumed for the typical UE, no matter
what the actual SINR value is. The definition of the ASE in (\ref{eq:ASE_def})
can better capture the dependence of the transmission rate on SINR
and maintains the monotonically decreasing feature of $A^{{\rm {ASE}}}\left(\lambda,\gamma_{0}\right)$
as $\gamma_{0}$ increases, but it is less tractable to analyze, as
it requires one more fold of numerical integral compared with~\cite{related_work_Jeff}.

Based on the definition of $p^{{\rm {cov}}}\left(\lambda,\gamma\right)$
in (\ref{eq:Coverage_Prob_def}), which is the complementary cumulative
distribution function (CCDF) of SINR, $f_{\mathit{\Gamma}}\left(\lambda,\gamma\right)$
can be expressed by
\begin{equation}
f_{\mathit{\Gamma}}\left(\lambda,\gamma\right)=\frac{\partial\left(1-p^{{\rm {cov}}}\left(\lambda,\gamma\right)\right)}{\partial\gamma},\label{eq:cond_SINR_PDF}
\end{equation}
where $p^{{\rm {cov}}}\left(\lambda,\gamma\right)$ is obtained from
Theorem~\ref{thm:p_cov_UAS1}.

\subsection{The ASE Crash Theorem\label{subsec:The-ASE-Crash}}

Considering the results of $p^{{\rm {cov}}}\left(\lambda,\gamma\right)$
and $A^{{\rm {ASE}}}\left(\lambda,\gamma_{0}\right)$ respectively
shown in (\ref{eq:Theorem_1_p_cov}) and (\ref{eq:ASE_def}), we propose
Theorem~\ref{thm:the_ASE_Falls_Theorem} to theoretically explain
the fundamental reasons of \emph{the ASE Crash} discussed in Section~\ref{sec:Introduction}.
\begin{thm}
{\small{}\label{thm:the_ASE_Falls_Theorem}}The ASE Crash Theorem:
If $L>0$ and $\gamma,\gamma_{0}<+\infty$, then $\underset{\lambda\rightarrow+\infty}{\lim}p^{{\rm {cov}}}\left(\lambda,\gamma\right)=0$
and $\underset{\lambda\rightarrow+\infty}{\lim}A^{{\rm {ASE}}}\left(\lambda,\gamma_{0}\right)=0$.
\end{thm}
\begin{IEEEproof}
See Appendix~C.
\end{IEEEproof}

In essence, Theorem~\ref{thm:the_ASE_Falls_Theorem} states that
when $\lambda$ is extremely large, e.g., in UDNs, both $p^{{\rm {cov}}}\left(\lambda,\gamma\right)$
and $A^{{\rm {ASE}}}\left(\lambda,\gamma_{0}\right)$ will decrease
towards zero with the network densification, and UEs will experience
service outage, thus creating \emph{the ASE Crash}. The fundamental
reason for this phenomena is revealed by the key point of the proof,
i.e., the signal-to-interference ratio (SIR) of the signal power and
any interference power at close proximity reaches one when $\lambda\rightarrow+\infty$.
In more detail, suppose that the 2D distance between the serving BS
and the UE and that between an arbitrary interfering BS and the UE
are denoted by $r$ and $\tau r,\left(1<\tau<+\infty\right)$, respectively.
In UDNs, we have $r\rightarrow0$ when $\lambda\rightarrow+\infty$.
Considering that $r\rightarrow0$ and $L$ is smaller than $d_{1}$
in practical SCNs~\cite{SCM_pathloss_model,TR36.828}, we can assume
that both the signal link and the interference link should be dominantly
characterized by \emph{the first-piece LoS path loss function} in
(\ref{eq:PL_BS2UE}), i.e., $\zeta_{1}^{{\rm {L}}}\left(w\right)=A_{1}^{{\rm {L}}}\left(\sqrt{r^{2}+L^{2}}\right)^{-\alpha_{1}^{{\rm {L}}}}$.
Thus, based on the 3D distances, we can obtain the SIR $\bar{\gamma}$
as
\begin{equation}
\bar{\gamma}=\frac{A_{1}^{{\rm {L}}}\left(\sqrt{r^{2}+L^{2}}\right)^{-\alpha_{1}^{{\rm {L}}}}}{A_{1}^{{\rm {L}}}\left(\sqrt{\tau^{2}r^{2}+L^{2}}\right)^{-\alpha_{1}^{{\rm {L}}}}}=\left(\sqrt{\frac{1}{1+\frac{\tau^{2}-1}{1+\frac{L^{2}}{r^{2}}}}}\right)^{-\alpha_{1}^{{\rm {L}}}}.\label{eq:SIR_toy_example}
\end{equation}
It is easy to show that $\bar{\gamma}\rightarrow1$ when $r\rightarrow0$
in UDNs. The intuition of such conclusion and why it leads to \emph{the
ASE Crash} are explained as follows,%

\begin{itemize}
\item Due to realistic BS deployments, a UE cannot be arbitrarily close
to it serving BS antenna. Thus, there exists a \emph{cap} for both
the signal power and the interference power. In our case, a UE cannot
be closer than $L$ to any BS due to a non-zero antenna height difference.
Such \emph{cap} can be seen in the numerator and the denominator of
$\bar{\gamma}$.
\item In the limit of ultra-dense networks where $\lambda\rightarrow+\infty$,
both the serving BS and the interfering BS at close proximity to the
typical UE will both be directly overhead and equidistant, i.e., $L$,
above the UE. As a result, the interfering link and the signal link
would almost have the same path loss, which leads to $\bar{\gamma}\rightarrow1$.
\item Consequently, the aggregate interference power will overwhelm the
signal power in UDNs due to the sheer number of strong interferers
growing from every direction around the typical UE.
\item It is very important to note that the above intuition of Theorem~\ref{thm:the_ASE_Falls_Theorem}
is rooted in the geometry of BS deployments and is valid regardless
of the path loss parameters and the multi-path fading model, e.g.,
Rician fading to be studied in the next subsection.
\end{itemize}

To sum up, in a UDN with conventional deployment (i.e., $L>0$), both
$p^{{\rm {cov}}}\left(\lambda,\gamma\right)$ and $A^{{\rm {ASE}}}\left(\lambda,\gamma_{0}\right)$
will plunge toward zero as $\lambda\rightarrow+\infty$, causing \emph{the
ASE Crash}. Its fundamental reason is \emph{the cap on the signal
power} because of the minimum signal-link distance tied to $L$, which
cannot be overcome with the densification. One way to avoid \emph{the
ASE Crash} is to remove the signal power cap by setting $L$ to zero,
which means lowering the BS antenna height, not just by a few meters,
but straight to the UE antenna height.

\subsection{The Analytical Results with Rician Fading\label{subsec:The-Analytical-Results-Rician}}

Note that Theorem~\ref{thm:p_cov_UAS1} is agnostic of the assumption
on multi-path fading and Lemma~\ref{lem:Pcov_LoS_and_NLoS_Rayleigh}
is derived for Rayleigh fading only, and thus it would be desirable
to extend Lemma~\ref{lem:Pcov_LoS_and_NLoS_Rayleigh} to the case
of Rician fading. Here, we consider a widely accepted model of i.i.d.
Rician fading~\cite{Yang2017denseSCNs}, where the ratio between
the power in the direct path and the power in the other scattered
paths is denoted by $K$. Note that we assume Rician fading and Rayleigh
fading for LoS and NLoS transmissions, respectively.

Based on such practical assumption, we derive Lemma~\ref{lem:Pcov_LoS_and_NLoS_Rician}
to replace Lemma~\ref{lem:Pcov_LoS_and_NLoS_Rayleigh} in the case
of Rician fading and Rayleigh fading for LoS and NLoS transmissions,
respectively.
\begin{algorithm*}
\begin{lem}
\begin{onehalfspace}
\label{lem:Pcov_LoS_and_NLoS_Rician}Considering Rician fading and
Rayleigh fading respectively for LoS transmissions and NLoS transmissions,
in Theorem~\ref{thm:p_cov_UAS1}, ${\rm {Pr}}\left[\frac{P\zeta_{n}^{{\rm {L}}}\left(\sqrt{r^{2}+L^{2}}\right)h}{I_{{\rm {agg}}}+P_{{\rm {N}}}}>\gamma\right]$
can be computed by
\begin{eqnarray}
{\rm {Pr}}\left[\frac{P\zeta_{n}^{{\rm {L}}}\left(\sqrt{r^{2}+L^{2}}\right)h}{I_{{\rm {agg}}}+P_{{\rm {N}}}}>\gamma\right]\hspace{-0.3cm} & = & \hspace{-0.3cm}\int_{-\infty}^{+\infty}\left(\frac{1-\exp\left(\frac{\textrm{j}\omega}{\gamma}\right)}{2\pi\textrm{j}\omega}\right)\mathcal{F}_{\frac{1}{SINR^{\textrm{L}}}}(\omega)d\omega,\label{eq:Pr_SINR_req_UAS1_LoS_Rician_lem}
\end{eqnarray}
where $\mathcal{F}_{\frac{1}{SINR^{\textrm{L}}}}(\omega)$ is written
as
\begin{eqnarray}
\mathcal{F}_{\frac{1}{SINR^{\textrm{L}}}}(\omega)\hspace{-0.3cm} & = & \hspace{-0.3cm}\int_{0}^{+\infty}ISR_{{\rm {aggL}}}^{{\rm {L}}}ISR_{{\rm {aggNL}}}^{{\rm {L}}}\exp\left(\textrm{j}\omega\frac{P_{{\rm {N}}}}{P\zeta_{n}^{{\rm {L}}}\left(\sqrt{r^{2}+L^{2}}\right)h}\right)f_{{\rm {fad}}}^{{\rm {L}}}\left(h\right)dh,\label{eq: final_char_func_UAS1_LoS_Rician_lem}
\end{eqnarray}
where
\begin{eqnarray}
\hspace{-0.1cm}\hspace{-0.1cm}\hspace{-0.1cm}\hspace{-0.1cm}\hspace{-0.1cm}\hspace{-0.1cm}ISR_{{\rm {aggL}}}^{{\rm {L}}}\hspace{-0.3cm} & = & \hspace{-0.3cm}\exp\left(-2\pi\lambda\int_{r}^{+\infty}\hspace{-0.1cm}\hspace{-0.1cm}\textrm{{\rm {Pr}}}^{{\rm {L}}}\left(\sqrt{u^{2}+L^{2}}\right)\hspace{-0.1cm}\left(1-\hspace{-0.1cm}\int_{0}^{+\infty}\hspace{-0.1cm}\hspace{-0.1cm}\hspace{-0.1cm}\exp\left(\textrm{j}\omega\frac{\zeta^{{\rm {L}}}\left(\sqrt{u^{2}+L^{2}}\right)x}{\zeta_{n}^{{\rm {L}}}\left(\sqrt{r^{2}+L^{2}}\right)h}\right)f_{{\rm {fad}}}^{{\rm {L}}}\left(x\right)dx\hspace{-0.1cm}\right)\hspace{-0.1cm}udu\right)\hspace{-0.1cm},\hspace{-0.1cm}\hspace{-0.1cm}\label{eq:ISR_LoS_interf_to_LoS_signal_lem}
\end{eqnarray}
\begin{eqnarray}
\hspace{-0.1cm}\hspace{-0.1cm}\hspace{-0.1cm}\hspace{-0.1cm}\hspace{-0.1cm}\hspace{-0.1cm}\hspace{-0.1cm}\hspace{-0.1cm}\hspace{-0.1cm}ISR_{{\rm {aggNL}}}^{{\rm {L}}}\hspace{-0.3cm} & = & \hspace{-0.3cm}\exp\hspace{-0.1cm}\left(\hspace{-0.1cm}-2\pi\lambda\hspace{-0.1cm}\int_{r_{1}}^{+\infty}\hspace{-0.1cm}\left[1-\textrm{{\rm {Pr}}}^{{\rm {L}}}\hspace{-0.1cm}\left(\sqrt{u^{2}+L^{2}}\right)\right]\hspace{-0.1cm}\left(\hspace{-0.1cm}1-\hspace{-0.1cm}\hspace{-0.1cm}\int_{0}^{+\infty}\hspace{-0.1cm}\hspace{-0.1cm}\hspace{-0.1cm}\hspace{-0.1cm}\exp\hspace{-0.1cm}\left(\textrm{j}\omega\frac{\zeta^{{\rm {NL}}}\left(\sqrt{u^{2}+L^{2}}\right)x}{\zeta_{n}^{{\rm {L}}}\left(\sqrt{r^{2}+L^{2}}\right)h}\right)\hspace{-0.1cm}f_{{\rm {fad}}}^{{\rm {NL}}}\left(x\right)\hspace{-0.1cm}dx\hspace{-0.1cm}\right)\hspace{-0.1cm}udu\hspace{-0.1cm}\right)\hspace{-0.1cm},\hspace{-0.1cm}\hspace{-0.1cm}\label{eq:ISR_NLoS_interf_to_LoS_signal_lem}
\end{eqnarray}
\begin{eqnarray}
f_{{\rm {fad}}}^{{\rm {NL}}}\left(x\right)\hspace{-0.3cm} & = & \hspace{-0.3cm}\exp\left(-x\right),\label{eq:PDF_Rayleigh_lem}
\end{eqnarray}
\begin{eqnarray}
f_{{\rm {fad}}}^{{\rm {L}}}\left(x\right)\hspace{-0.3cm} & = & \hspace{-0.3cm}\left(K+1\right)\exp\left(-K-\left(K+1\right)x\right)I_{0}\left(2\sqrt{K\left(K+1\right)x}\right),\label{eq:PDF_Rician_lem}
\end{eqnarray}
where $I_{0}\left(\cdot\right)$ is the 0-th order modified Bessel
function of the first kind~\cite{Book_Integrals}.

$\quad$Furthermore, ${\rm {Pr}}\left[\frac{P\zeta_{n}^{{\rm {NL}}}\left(\sqrt{r^{2}+L^{2}}\right)h}{I_{{\rm {agg}}}+P_{{\rm {N}}}}>\gamma\right]$
can be computed by
\begin{eqnarray}
{\rm {Pr}}\left[\frac{P\zeta_{n}^{{\rm {NL}}}\left(\sqrt{r^{2}+L^{2}}\right)h}{I_{{\rm {agg}}}+P_{{\rm {N}}}}>\gamma\right]\hspace{-0.3cm} & = & \hspace{-0.3cm}\int_{-\infty}^{+\infty}\left(\frac{1-\exp\left(\frac{\textrm{j}\omega}{\gamma}\right)}{2\pi\textrm{j}\omega}\right)\mathcal{F}_{\frac{1}{SINR^{\textrm{NL}}}}(\omega)d\omega,\label{eq:Pr_SINR_req_UAS1_NLoS_Rayleigh_lem}
\end{eqnarray}
where $\mathcal{F}_{\frac{1}{SINR^{\textrm{NL}}}}(\omega)$ is written
as
\begin{eqnarray}
\mathcal{F}_{\frac{1}{SINR^{\textrm{NL}}}}(\omega)\hspace{-0.3cm} & = & \hspace{-0.3cm}\int_{0}^{+\infty}ISR_{{\rm {aggL}}}^{{\rm {NL}}}ISR_{{\rm {aggNL}}}^{{\rm {NL}}}\exp\left(\textrm{j}\omega\frac{P_{{\rm {N}}}}{P\zeta_{n}^{{\rm {NL}}}\left(\sqrt{r^{2}+L^{2}}\right)h}\right)f_{{\rm {fad}}}^{{\rm {NL}}}\left(h\right)dh,\label{eq: final_char_func_UAS1_NLoS_Rayleigh_lem}
\end{eqnarray}
where
\begin{eqnarray}
\hspace{-0.1cm}\hspace{-0.1cm}\hspace{-0.1cm}\hspace{-0.1cm}\hspace{-0.1cm}\hspace{-0.1cm}ISR_{{\rm {aggL}}}^{{\rm {NL}}}\hspace{-0.3cm} & = & \hspace{-0.3cm}\exp\left(-2\pi\lambda\int_{_{r_{2}}}^{+\infty}\hspace{-0.1cm}\hspace{-0.1cm}\textrm{{\rm {Pr}}}^{{\rm {L}}}\left(\sqrt{u^{2}+L^{2}}\right)\hspace{-0.1cm}\left(1-\hspace{-0.1cm}\int_{0}^{+\infty}\hspace{-0.1cm}\hspace{-0.1cm}\hspace{-0.1cm}\exp\left(\textrm{j}\omega\frac{\zeta^{{\rm {L}}}\left(\sqrt{u^{2}+L^{2}}\right)x}{\zeta_{n}^{{\rm {NL}}}\left(\sqrt{r^{2}+L^{2}}\right)h}\right)f_{{\rm {fad}}}^{{\rm {L}}}\left(x\right)dx\hspace{-0.1cm}\right)\hspace{-0.1cm}udu\right)\hspace{-0.1cm},\hspace{-0.1cm}\hspace{-0.1cm}\label{eq:ISR_LoS_interf_to_NLoS_signal_lem}
\end{eqnarray}
\begin{eqnarray}
\hspace{-0.1cm}\hspace{-0.1cm}\hspace{-0.1cm}\hspace{-0.1cm}\hspace{-0.1cm}\hspace{-0.1cm}\hspace{-0.1cm}\hspace{-0.1cm}\hspace{-0.1cm}ISR_{{\rm {aggNL}}}^{{\rm {NL}}}\hspace{-0.3cm} & = & \hspace{-0.3cm}\exp\hspace{-0.1cm}\left(\hspace{-0.1cm}-2\pi\lambda\hspace{-0.1cm}\int_{r}^{+\infty}\hspace{-0.1cm}\left[1-\textrm{{\rm {Pr}}}^{{\rm {L}}}\hspace{-0.1cm}\left(\sqrt{u^{2}+L^{2}}\right)\right]\hspace{-0.1cm}\left(\hspace{-0.1cm}1-\hspace{-0.1cm}\hspace{-0.1cm}\int_{0}^{+\infty}\hspace{-0.1cm}\hspace{-0.1cm}\hspace{-0.1cm}\hspace{-0.1cm}\exp\hspace{-0.1cm}\left(\textrm{j}\omega\frac{\zeta^{{\rm {NL}}}\left(\sqrt{u^{2}+L^{2}}\right)x}{\zeta_{n}^{{\rm {NL}}}\left(\sqrt{r^{2}+L^{2}}\right)h}\right)\hspace{-0.1cm}f_{{\rm {fad}}}^{{\rm {NL}}}\left(x\right)\hspace{-0.1cm}dx\hspace{-0.1cm}\right)\hspace{-0.1cm}udu\hspace{-0.1cm}\right)\hspace{-0.1cm},\hspace{-0.1cm}\hspace{-0.1cm}\label{eq:ISR_NLoS_interf_to_NLoS_signal_lem}
\end{eqnarray}
\end{onehalfspace}
\end{lem}
\begin{IEEEproof}
\begin{onehalfspace}
See Appendix~D.
\end{onehalfspace}
\end{IEEEproof}
\end{algorithm*}
 Note that Lemma~\ref{lem:Pcov_LoS_and_NLoS_Rician} is derived following
a similar logic as that in our previous work~\cite{Yang2017denseSCNs}.
Moreover, Lemma~\ref{lem:Pcov_LoS_and_NLoS_Rician} can be easily
extended to other multi-path fading for LoS transmissions by using
alternative PDF functions of $f_{{\rm {fad}}}^{{\rm {L}}}\left(h\right)$.
It should also be noted that a more practical model of Rician fading
can be found in~\cite{SCM_pathloss_model}, where the $K$ factor
in dB scale is modeled as $K^{\textrm{[dB]}}=13-0.03w$, where $w$
is the 3D distance in meter. The corresponding theoretical analysis
would be more challenging than Lemma~\ref{lem:Pcov_LoS_and_NLoS_Rician}
because such distance-dependent Rician fading is not i.i.d. for all
links. Thus, it is left as our future work.

\subsection{The 3GPP Special Cases\label{subsec:The-3GPP-Special-Cases}}

Regarding realistic path loss models, we consider a two-piece path
loss and a linear LoS probability functions defined by the 3GPP~\cite{TR36.828,SCM_pathloss_model}.
Specifically, we use the path loss function $\zeta\left(w\right)$,
defined in the 3GPP as~\cite{TR36.828}
\begin{equation}
\zeta\left(w\right)=\begin{cases}
\begin{array}{l}
A^{{\rm {L}}}w^{-\alpha^{{\rm {L}}}},\\
A^{{\rm {NL}}}w^{-\alpha^{{\rm {NL}}}},
\end{array}\hspace{-0.3cm} & \begin{array}{l}
\textrm{\textrm{LoS:}~}\textrm{Pr}^{\textrm{L}}\left(w\right)\\
\textrm{\textrm{NLoS:}~}1-\textrm{Pr}^{\textrm{L}}\left(w\right)
\end{array}\end{cases},\label{eq:PL_BS2UE_2slopes}
\end{equation}
together with a linear LoS probability function of $\textrm{Pr}^{\textrm{L}}\left(r\right)$,
defined in the 3GPP as~\cite{SCM_pathloss_model}
\begin{equation}
\textrm{Pr}^{\textrm{L}}\left(w\right)=\begin{cases}
\begin{array}{l}
1-\frac{w}{d_{1}},\\
0,
\end{array}\hspace{-0.3cm} & \begin{array}{l}
L<w\leq d_{1}\\
w>d_{1}
\end{array}\end{cases},\label{eq:LoS_Prob_func_linear}
\end{equation}
where $d_{1}$ is a constant~\cite{SCM_pathloss_model}.

Considering the general path loss model presented in (\ref{eq:prop_PL_model}),
the combined path loss model presented in (\ref{eq:PL_BS2UE_2slopes})
and (\ref{eq:LoS_Prob_func_linear}) can be deemed as a special case
of (\ref{eq:prop_PL_model}) with the following substitution: $N=2$,
$\zeta_{1}^{\textrm{L}}\left(w\right)=\zeta_{2}^{\textrm{L}}\left(w\right)=A^{{\rm {L}}}w^{-\alpha^{{\rm {L}}}}$,
$\zeta_{1}^{\textrm{NL}}\left(w\right)=\zeta_{2}^{\textrm{NL}}\left(w\right)=A^{{\rm {NL}}}w^{-\alpha^{{\rm {NL}}}}$,
$\textrm{Pr}_{1}^{\textrm{L}}\left(w\right)=1-\frac{w}{d_{1}}$, and
$\textrm{Pr}_{2}^{\textrm{L}}\left(w\right)=0$. For clarity, this
3GPP special case is referred to as \textbf{3GPP Case~1} in the sequel.%
{}

To demonstrate that our conclusions have general significance, we
consider another widely used LoS probability function, which is an
exponential function defined in the 3GPP as~\cite{TR36.828}
\begin{equation}
\textrm{Pr}^{\textrm{L}}\left(r\right)=\begin{cases}
\begin{array}{l}
1-5\exp\left(-R_{1}/r\right),\\
5\exp\left(-r/R_{2}\right),
\end{array} & \begin{array}{l}
L<r\leq d_{1}\\
r>d_{1}
\end{array}\end{cases},\label{eq:LoS_Prob_func_reverseS_shape}
\end{equation}
where $R_{1}$ and $R_{2}$ are constants, and $d_{1}=\frac{R_{1}}{\ln10}$.
The combination of the path loss function in (\ref{eq:PL_BS2UE_2slopes})
and the LoS probability function in (\ref{eq:LoS_Prob_func_reverseS_shape})
can then be deemed as a special case of the proposed path loss model
in (\ref{eq:prop_PL_model}) with the following substitution: $N=2$,
$\zeta_{1}^{\textrm{L}}\left(r\right)=\zeta_{2}^{\textrm{L}}\left(r\right)=A^{{\rm {L}}}r^{-\alpha^{{\rm {L}}}}$,
$\zeta_{1}^{\textrm{NL}}\left(r\right)=\zeta_{2}^{\textrm{NL}}\left(r\right)=A^{{\rm {NL}}}r^{-\alpha^{{\rm {NL}}}}$,
$\textrm{Pr}_{1}^{\textrm{L}}\left(r\right)=1-5\exp\left(-R_{1}/r\right)$,
and $\textrm{Pr}_{2}^{\textrm{L}}\left(r\right)=5\exp\left(-r/R_{2}\right)$.
For clarity, this combined case with both the path loss function and
the LoS probability function coming from~\cite{TR36.828} is referred
to as \textbf{3GPP Case~2} hereafter.

As justified in~\cite{our_work_TWC2016}, we mainly use 3GPP Case~1
to generate the numerical results in Section~\ref{sec:Simulation-and-Discussion},
because it provides tractable results for $\left\{ f_{R,n}^{Path}\left(r\right)\right\} $
and $\left\{ \mathscr{L}_{I_{{\rm {agg}}}}^{Path}\left(s\right)\right\} $
in%
{} Theorem~\ref{thm:p_cov_UAS1}. To further improve the tractability
of the results for 3GPP Case~1, we further propose an approximation
technique in Appendix~E. Nevertheless, we will numerically investigate
3GPP Case~2 using Theorem~\ref{thm:p_cov_UAS1} in Section~\ref{sec:Simulation-and-Discussion},
and we will show that similar conclusions like those for 3GPP Case~1
can also be drawn for 3GPP Case~2.%

Also note that eventually we may need to conduct hardware experiments
in real-world fields to verify the existence of \emph{the ASE Crash}.
However, since there has been no UDN deployed in the world yet, it
is difficult to do so at the current stage. Therefore, we use the
propagation models that have been widely accepted by most industrial
companies in the 3GPP~\cite{TR36.828,SCM_pathloss_model} to obtain
simulation results to verify our theoretical findings. Due to the
limitation in using 3GPP Cases~1 and~2 to represent the real-world
environment, the quantitative results in our study might deviate from
those measured in practice, and thus we should focus more on the qualitative
conclusions of our theoretical discoveries.

\section{Simulation and Discussion\label{sec:Simulation-and-Discussion}}

In this section, we investigate the network performance and use numerical
results to establish the accuracy of our analysis. According to Tables
A.1-3, A.1-4 and A.1-7 of~\cite{TR36.828} and~\cite{SCM_pathloss_model},
we adopt the following parameters for 3GPP Case~1: $d_{1}=300$\ m,
$\alpha^{{\rm {L}}}=2.09$, $\alpha^{{\rm {NL}}}=3.75$, $A^{{\rm {L}}}=10^{-10.38}$,
$A^{{\rm {NL}}}=10^{-14.54}$, $P=24$\ dBm, $P_{{\rm {N}}}=-95$\ dBm
(including a noise figure of 9\ dB at the UE). We have also investigated
the results for a single-slope path loss model shown in (\ref{eq:single_slope_PL_BS2UE})~\cite{Jeff2011},
where only one path loss exponent $\alpha$ is defined, the value
of which is assumed to be $\alpha=3.75$.%

\subsection{Validation of Theorem~\ref{thm:p_cov_UAS1} on the Coverage Probability\label{subsec:Sim-p-cov-3GPP-Case-1}}

\begin{figure}
\noindent \centering{}\includegraphics[width=8cm]{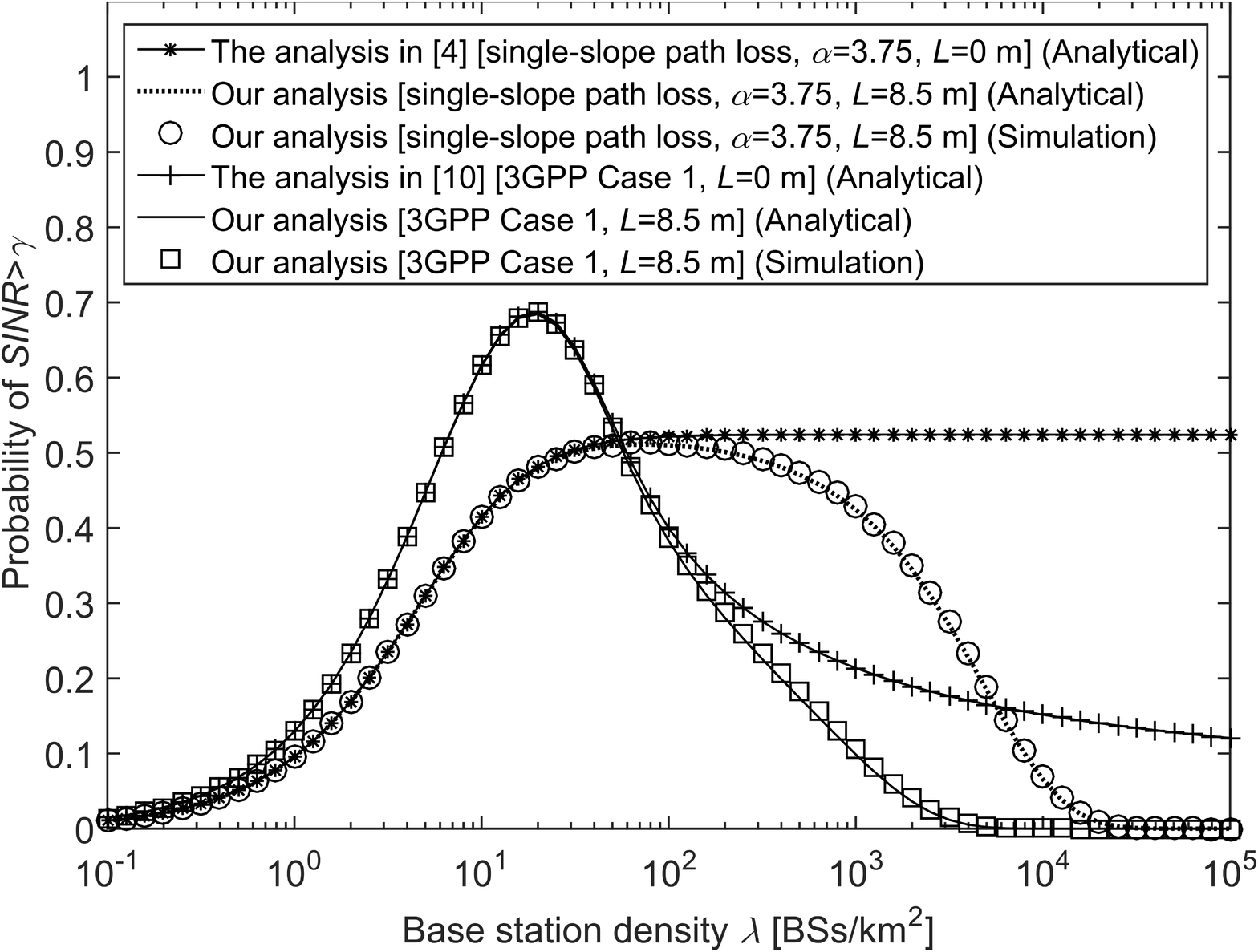}\renewcommand{\figurename}{Fig.}\caption{\label{fig:p_cov_linear_3Dmod_fixedPower24dBm_gamma0dB_wBaseline}$p^{{\rm {cov}}}\left(\lambda,\gamma\right)$
vs. $\lambda$ with $\gamma=0\,\textrm{dB}$ (Rayleigh fading only).}
\end{figure}

In Fig.~\ref{fig:p_cov_linear_3Dmod_fixedPower24dBm_gamma0dB_wBaseline},
we show the results of $p^{{\rm {cov}}}\left(\lambda,\gamma\right)$
with $\gamma=0\,\textrm{dB}$. Regarding the non-zero value of $L$,
as explained in Section~\ref{sec:Introduction}, the BS antenna and
the UE antenna heights are set to 10\,m and 1.5\,m, respectively~\cite{TR36.814},
thus $L=\left|10-1.5\right|=8.5$\,m.

As can be observed from Fig.~\ref{fig:p_cov_linear_3Dmod_fixedPower24dBm_gamma0dB_wBaseline},
our analytical results given by Theorem~\ref{thm:p_cov_UAS1} and
Lemma~\ref{lem:Pcov_LoS_and_NLoS_Rayleigh} match the simulation
results very well, which validates the accuracy of our analysis. From
Fig.~\ref{fig:p_cov_linear_3Dmod_fixedPower24dBm_gamma0dB_wBaseline},
we can draw the following observations, which confirm our discussion
in Section~\ref{sec:Introduction}:
\begin{itemize}
\item For the single-slope path loss model with $L=0$\,m, \emph{the BS
density does NOT matter}, since the coverage probability approaches
a constant for UDNs~\cite{Jeff2011}, e.g., $\lambda>10^{2}\,\textrm{BSs/km}^{2}$.
\item For the 3GPP Case~1 path loss model with $L=0$\,m, \emph{the BS
density DOES matter,} since that coverage probability will decrease
as $\lambda$ increases when the network is dense enough, e.g., $\lambda>20\,\textrm{BSs/km}^{2}$,
due to the transition of a large number of interference paths from
NLoS to LoS~\cite{our_work_TWC2016}.%
\item For both the 3GPP Case~1 and the single-slope path loss model shown
in (\ref{eq:single_slope_PL_BS2UE}), when $L=8.5$\,m, the coverage
probability shows a determined trajectory toward zero in the UDN regime%
{} as explained in Theorem~\ref{thm:the_ASE_Falls_Theorem}.
\end{itemize}

\subsection{Validation of Theorem~\ref{thm:the_ASE_Falls_Theorem} on the ASE
Crash\label{subsec:Sim-ASE-3GPP-Case-1}}

\begin{figure}
\noindent \centering{}\includegraphics[width=8cm]{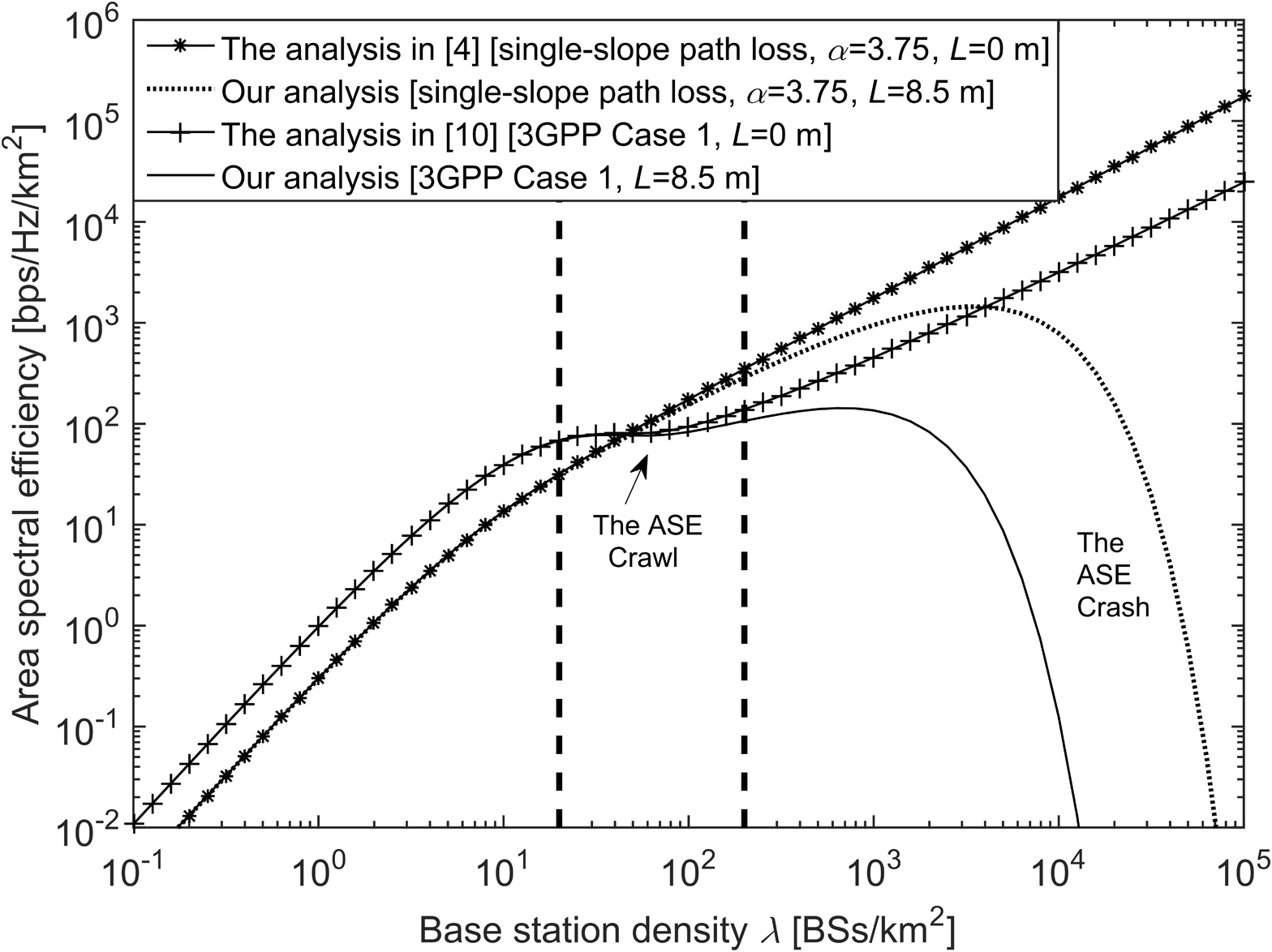}\renewcommand{\figurename}{Fig.}\caption{\label{fig:ASE_linear_3Dmod_fixedPower24dBm_gamma0dB_wBaseline}$A^{{\rm {ASE}}}\left(\lambda,\gamma_{0}\right)$
vs. $\lambda$ with $\gamma_{0}=0\,\textrm{dB}$ (Rayleigh fading
only).}
\end{figure}

In Fig.~\ref{fig:ASE_linear_3Dmod_fixedPower24dBm_gamma0dB_wBaseline},
we show the results of $A^{{\rm {ASE}}}\left(\lambda,\gamma_{0}\right)$
with $\gamma_{0}=0\,\textrm{dB}$.
 Due to the significant accuracy of our analysis on $p^{{\rm {cov}}}\left(\lambda,\gamma\right)$
demonstrated in Fig.~\ref{fig:p_cov_linear_3Dmod_fixedPower24dBm_gamma0dB_wBaseline},
we only show analytical results of $A^{{\rm {ASE}}}\left(\lambda,\gamma_{0}\right)$
in Fig.~\ref{fig:ASE_linear_3Dmod_fixedPower24dBm_gamma0dB_wBaseline},
because $A^{{\rm {ASE}}}\left(\lambda,\gamma_{0}\right)$ is computed
from $p^{{\rm {cov}}}\left(\lambda,\gamma\right)$ as discussed in
Subsection~\ref{subsec:The-ASE-Performance}.

Fig.~\ref{fig:ASE_linear_3Dmod_fixedPower24dBm_gamma0dB_wBaseline}
is essentially the same as Fig.~\ref{fig:comp_ASE_2Gto5G} with the
same marker styles, except that the results for the simplistic single-slope
path loss model with $L=8.5$\,m are also plotted. From Fig.~\ref{fig:ASE_linear_3Dmod_fixedPower24dBm_gamma0dB_wBaseline},
we can confirm the key observations presented in Section~\ref{sec:Introduction}:
\begin{itemize}
\item For the single-slope path loss model with $L=0$\,m, the ASE performance
scales linearly with $\lambda$~\cite{Jeff2011}. The result is promising,
but it might not be the truth in reality.
\item For the 3GPP Case~1 path loss model with $L=0$\,m, the ASE suffers
from a slow growth or even a small\emph{ decrease} when $\lambda\in\left[20,200\right]\,\textrm{BSs/km}^{2}$,
i.e., \emph{the ASE Crawl}~\cite{our_work_TWC2016}. After \emph{the
ASE Crawl}, the ASE grows almost linearly again as the network further
evolves to a UDN, e.g., $\lambda>10^{3}\thinspace\textrm{BSs/km}^{2}$~\cite{our_work_TWC2016}.
\item For both path loss models with $L=8.5$\,m, the ASE suffers from
severe performance loss in UDNs due to \emph{the ASE Crash}, as explained
in Theorem~\ref{thm:the_ASE_Falls_Theorem}.
\item Here, we have established a baseline ASE Crash performance with the
assumptions of \textquotedblleft 3GPP Case~1, $L=8.5$\,m, Rayleigh
fading only\textquotedblright . In the following subsections, all
of the numerical results will be \emph{compared against such baseline
ASE Crash performance} to show the performance impacts of various
factors.
\end{itemize}

\subsection{The Performance Impact of $L$ on the ASE Crash\label{subsec:impact_of_L_on_ASE}}

\begin{figure}
\noindent \centering{}\includegraphics[width=8cm]{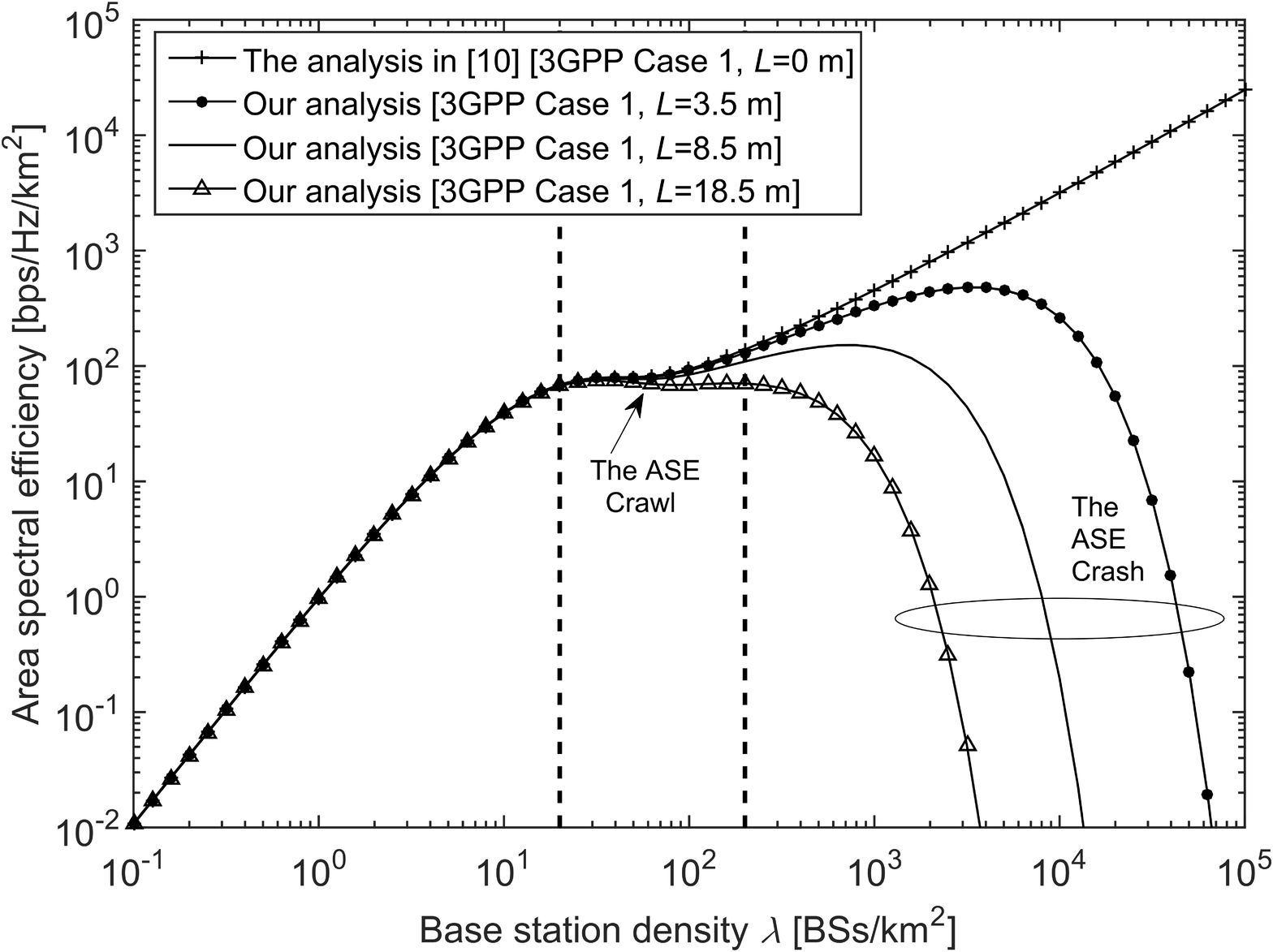}\renewcommand{\figurename}{Fig.}\caption{\label{fig:ASE_3Dmod_fixedPower24dBm_gamma0dB_variousL}$A^{{\rm {ASE}}}\left(\lambda,\gamma_{0}\right)$
vs. $\lambda$ with $\gamma_{0}=0\,\textrm{dB}$ and various $L$
(Rayleigh fading only).}
\end{figure}

In Fig.~\ref{fig:ASE_3Dmod_fixedPower24dBm_gamma0dB_variousL}, we
show the results of $A^{{\rm {ASE}}}\left(\lambda,\gamma_{0}\right)$
with $\gamma_{0}=0\,\textrm{dB}$ and various values of $L$. We assume
that the UE antenna height is still 1.5\,m, but the BS antenna height
changes to 5\,m, 10\,m and 20\,m, respectively.
 Accordingly, $L$ takes the values of 3.5\,m, 8.5\,m and 18.5\,m,
respectively. Our key conclusions are summarized in the following:
\begin{itemize}
\item The larger the $L$, the severer \emph{the ASE Crash}. This is because
a larger $L$ implies a tighter cap on the signal power and the interference
power, which leads to an earlier arrival of $\bar{\gamma}=1$ in (\ref{eq:SIR_toy_example})
and thus \emph{the ASE Crash}.
\item Compared with the baseline ASE Crash performance, the reduction of
$L$ from $8.5$\,m to $3.5$\,m can delay \emph{the ASE Crash}
from around $\lambda=10^{4}\thinspace\textrm{BSs/km}^{2}$ to around
$\lambda=4\times10^{4}\thinspace\textrm{BSs/km}^{2}$ when the ASE
hits 1\,$\textrm{bps/Hz/km}^{2}$. However, it is important to note
that the ASE with $L=3.5$\,m peaks at around $\lambda^{*}=3000\thinspace\textrm{BSs/km}^{2}$,
but it still suffers from a 60\,\% loss compared with that with $L=0$\,m
at $\lambda^{*}$.
\end{itemize}

\subsection{The Performance Impact of Antenna Pattern and Downtilt on the ASE
Crash\label{subsec:downtilt_impact_on_ASE}}

As discussed in Section~\ref{sec:System-Model}, via downtilt in
the vertical domain, a practical antenna can target its antenna beam
towards a given direction, which may affect \emph{the ASE Crash} behavior.
Here, we adopt the antenna pattern and downtilt model proposed in~\cite{Li2015antenna}.
More specifically, in our analysis, the path loss function $\zeta\left(w\right)$
in (\ref{eq:prop_PL_model}) should be replaced by $\zeta\left(w\right)10^{\frac{1}{10}G\left(\varphi,\theta,\theta_{\textrm{tilt}}\right)}$,
where $G\left(\varphi,\theta,\theta_{\textrm{tilt}}\right)$ is the
antenna gain in the dB unit and it can be expressed by
\begin{equation}
G\left(\varphi,\theta,\theta_{\textrm{tilt}}\right)=G_{\textrm{M}}+G_{\textrm{H}}\left(\varphi\right)+G_{\textrm{V}}\left(\theta,\theta_{\textrm{tilt}}\right),\label{eq:antenna_gain_dB}
\end{equation}
where $\varphi$ and $\theta$ are the angles of arrival in the horizontal
and vertical planes, respectively, $\theta_{\textrm{tilt}}$ is the
electrical downtilt angle of the vertical antenna beam, $G_{\textrm{M}}$
is the maximum antenna gain in dB, $G_{\textrm{H}}\left(\varphi\right)$
is the horizontal attenuation offset in dB, and $G_{\textrm{V}}\left(\theta,\theta_{\textrm{tilt}}\right)$
is the vertical attenuation offset in dB.

Considering a 4-element half-wave dipole antenna, we have $G_{\textrm{M}}=8.15\,\textrm{dB}$~\cite{Li2015antenna}.
For the horizontal pattern, as discussed in Section~\ref{sec:System-Model},
we consider a omni-directional antenna, i.e., $G_{\textrm{H}}\left(\varphi\right)=0\,\textrm{dB}$.
For the vertical pattern, $G_{\textrm{V}}\left(\theta,\theta_{\textrm{tilt}}\right)$
is formulated according to~\cite{Li2015antenna} as
\begin{equation}
G_{\textrm{V}}\left(\theta,\theta_{\textrm{tilt}}\right)=\max\left\{ 10\log_{10}\left|\cos^{n}\left(\theta-\theta_{\textrm{tilt}}\right)\right|,F_{\textrm{V}}\right\} ,\label{eq:vertical_pattern}
\end{equation}
where $n$ equals to 47.64 for a 4-element half-wave dipole antenna
with a vertical half-power band-width (HPBW) of $B_{\textrm{V}}=19.5$$\,$degrees,
and $F_{\textrm{V}}$ is the vertical side-lobe level (SLL), which
is set to -12$\,$dB in~\cite{Li2015antenna}. As discussed in Section~\ref{sec:System-Model},
it is important to note that in practice $\theta_{\textrm{tilt}}$
becomes larger as the BS density $\lambda$ increases. According to~\cite{Fischer2002tiltAngle},
$\theta_{\textrm{tilt}}$ can be empirically modeled as
\begin{equation}
\theta_{\textrm{tilt}}=\arctan\left(\frac{L}{r^{\textrm{cov}}}\right)+zB_{\textrm{V}},\label{eq:tilt_angle}
\end{equation}
where $r^{\textrm{cov}}$ is the average distance from a cell-edge
UE to its serving BS given by $r^{\textrm{cov}}=\sqrt{\frac{1}{\lambda\pi}}$
in our analysis, and $z$ is an empirical parameter achieving a good
trade-off between the received signal power and the resulting inter-cell
interference. In~\cite{Fischer2002tiltAngle}, $z$ is set to 0.7.

\begin{figure}
\noindent \centering{}\includegraphics[width=7cm]{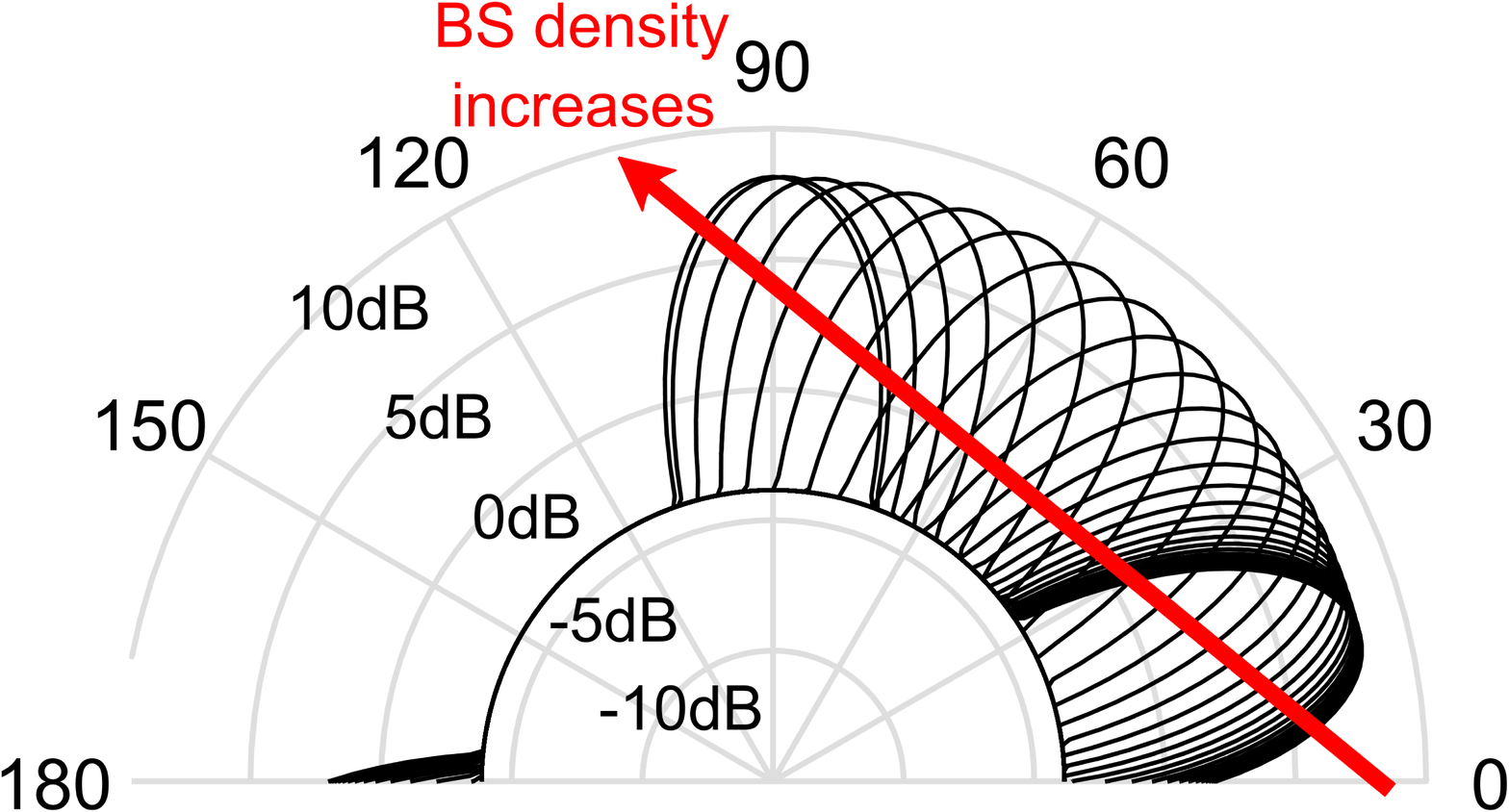}\renewcommand{\figurename}{Fig.}\caption{\label{fig:antenna_gain_dB}$G\left(\varphi,\theta,\theta_{\textrm{tilt}}\right)$
vs. $\theta$ for a 4-element half-wave dipole antenna~\cite{Li2015antenna,Fischer2002tiltAngle}.}
\end{figure}

Plugging (\ref{eq:tilt_angle}) into (\ref{eq:vertical_pattern}),
we can obtain the antenna gain $G\left(\varphi,\theta,\theta_{\textrm{tilt}}\right)$
considering practical antenna pattern and downtilt. Such results are
illustrated in Fig.~\ref{fig:antenna_gain_dB}.
 From this figure, we can observe that the downtilt angle of the vertical
antenna beam gradually increases from around 10 degrees to 90 degrees
as the network densifies, and the maximum antenna gain is $G_{\textrm{M}}=8.15\,\textrm{dB}$
at the direction of such downtilt angle.

\begin{figure}
\noindent \centering{}\includegraphics[width=8cm]{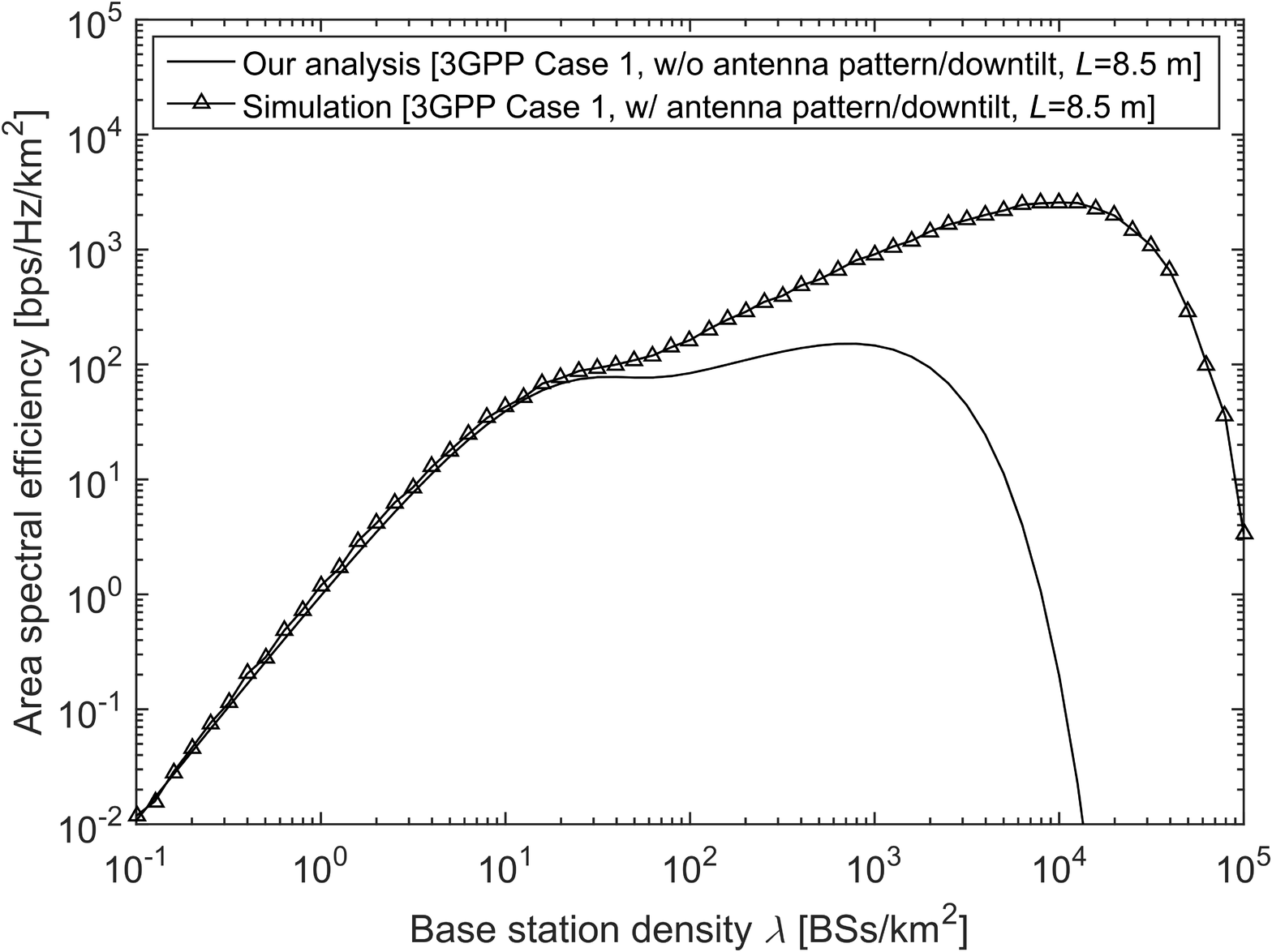}\renewcommand{\figurename}{Fig.}\caption{\label{fig:ASE_3Dmod_fixedPower24dBm_gamma0dB_downtilt}$A^{{\rm {ASE}}}\left(\lambda,\gamma_{0}\right)$
vs. $\lambda$ with $\gamma_{0}=0\,\textrm{dB}$ and the practical
antenna pattern and downtilt shown in Fig.~\ref{fig:antenna_gain_dB}
(Rayleigh fading only).}
\end{figure}

Based on the results of $G\left(\varphi,\theta,\theta_{\textrm{tilt}}\right)$
displayed in Fig.~\ref{fig:antenna_gain_dB}, we investigate the
performance of $A^{{\rm {ASE}}}\left(\lambda,\gamma_{0}\right)$ with
practical antenna pattern and downtilt in Fig.~\ref{fig:ASE_3Dmod_fixedPower24dBm_gamma0dB_downtilt}.
 Our key conclusions from Fig.~\ref{fig:ASE_3Dmod_fixedPower24dBm_gamma0dB_downtilt}
are drawn as follows:
\begin{itemize}
\item The practical antenna pattern and downtilt shown in Fig.~\ref{fig:antenna_gain_dB}
help to alleviate \emph{the ASE Crash} because they constrain the
BS energy emission within certain geometrical areas. However, \emph{the
ASE Crash} still emerges in UDNs because the cap on the signal power
persists, even if the BS antenna faces downward with a downtilt angle
of 90 degrees.
\item Compared with the baseline ASE Crash performance, the practical antenna
pattern and downtilt can delay \emph{the ASE Crash} from around $\lambda=10^{4}\thinspace\textrm{BSs/km}^{2}$
to more than $\lambda=10^{5}\thinspace\textrm{BSs/km}^{2}$ when the
ASE declines to 1\,$\textrm{bps/Hz/km}^{2}$.
\end{itemize}

\subsection{The Performance Impact of Rician Fading on the ASE Crash\label{subsec:Rician_impact_on_ASE}}

\begin{figure}
\noindent \centering{}\includegraphics[width=8cm]{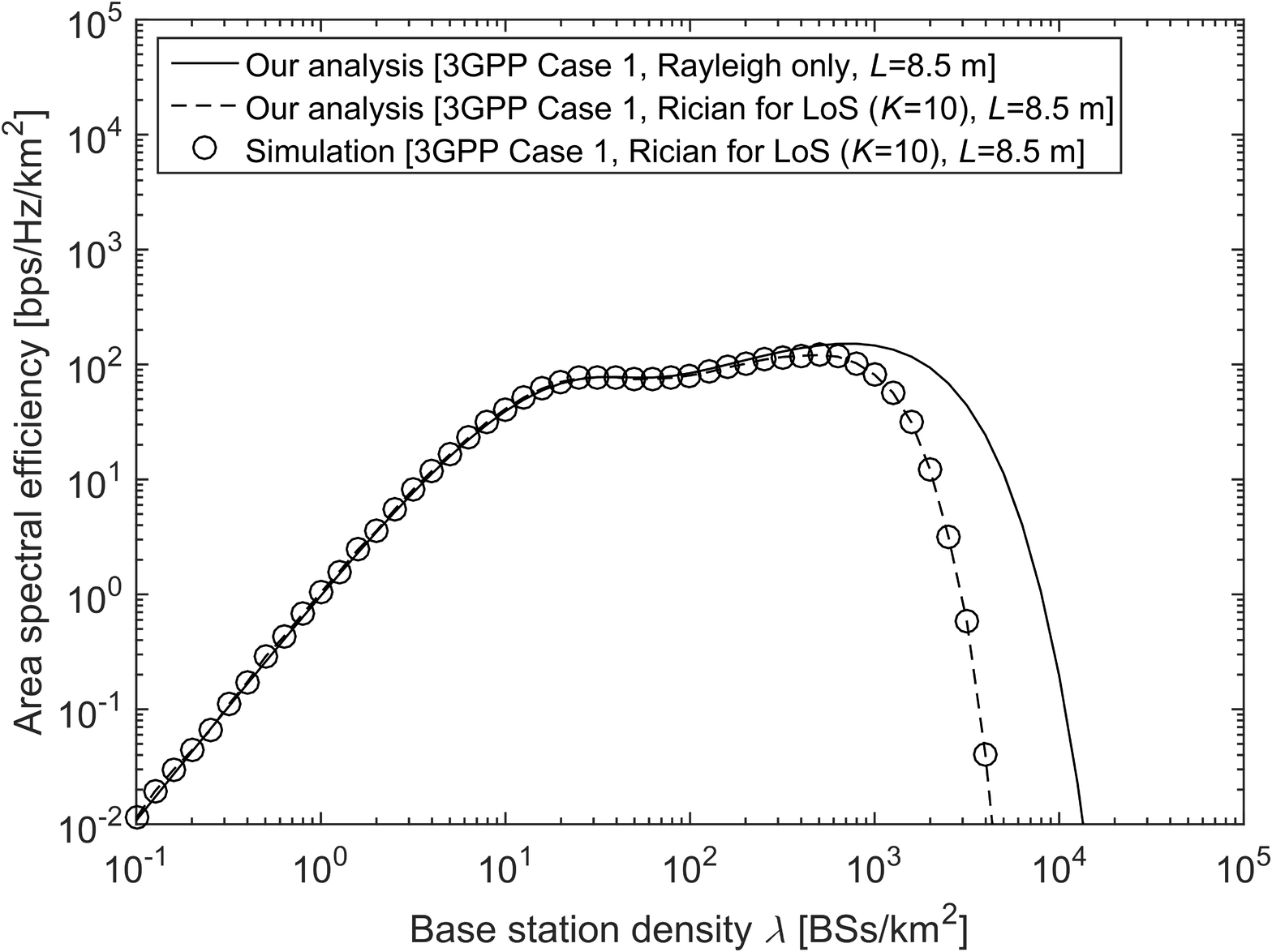}\renewcommand{\figurename}{Fig.}\caption{\label{fig:ASE_3Dmod_fixedPower24dBm_gamma0dB_Rician}$A^{{\rm {ASE}}}\left(\lambda,\gamma_{0}\right)$
vs. $\lambda$ with $\gamma_{0}=0\,\textrm{dB}$ (Rayleigh fading
for NLoS transmissions and Rician fading ($K=10$) for LoS transmissions).}
\end{figure}

In Fig.~\ref{fig:ASE_3Dmod_fixedPower24dBm_gamma0dB_Rician}, we
investigate the performance of $A^{{\rm {ASE}}}\left(\lambda,\gamma_{0}\right)$
under the assumptions of Rayleigh fading for NLoS transmissions and
Rician fading ($K=10$) for LoS transmissions.
 The analytical results are obtained from Theorem~\ref{thm:p_cov_UAS1}
and Lemma~\ref{lem:Pcov_LoS_and_NLoS_Rician} presented in Subsection~\ref{subsec:The-Analytical-Results-Rician}.%
{} Fig.~\ref{fig:ASE_3Dmod_fixedPower24dBm_gamma0dB_Rician} shows
that our analytical results match the simulation results very well,
which validates the accuracy of our analysis. Our key conclusions
from Fig.~\ref{fig:ASE_3Dmod_fixedPower24dBm_gamma0dB_Rician} are
summarized as follows:%

\begin{itemize}
\item \textcolor{black}{We can see that Rician fading makes }\emph{the ASE
Crash}\textcolor{black}{{} worse, which }takes effect earlier than the
case with Rayleigh fading. The intuition is that \textcolor{black}{the
randomness in channel fluctuation associated with Rician fading is
much weaker than that associated with Rayleigh fading due to the large
}$K$ factor in UDNs~\cite{SCM_pathloss_model}. With \textcolor{black}{Rayleigh
fading, some UE in outage might be opportunistically saved by favorable
channel fluctuation of the signal power, while with Rician fading,
such outage case becomes more deterministic due to lack of channel
variation, thus leading to a severer }\emph{ASE Crash}.\textcolor{black}{}%
\item Compared with the baseline ASE Crash performance, \textcolor{black}{the
investigated Rician fading} will bring forward \emph{the ASE Crash}
from around $\lambda=10^{4}\thinspace\textrm{BSs/km}^{2}$ to $\lambda=3\times10^{3}\thinspace\textrm{BSs/km}^{2}$
when the ASE is merely 1\,$\textrm{bps/Hz/km}^{2}$.
\end{itemize}

\subsection{The Performance Impact of 3GPP Case~2 on the ASE Crash\label{subsec:impact_of_3GPP_Case2_on_ASE}}

In this subsection, we investigate the ASE performance for 3GPP Case~2,
which has been discussed in Subsection~\ref{subsec:The-3GPP-Special-Cases}.
The parameters in the LoS probability function $\textrm{Pr}^{\textrm{L}}\left(w\right)$
of 3GPP Case~2 are set to $R_{1}=156$\ m and $R_{2}=30$\ m~\cite{TR36.828}.
First, we directly apply the numerical integration in Theorem~\ref{thm:p_cov_UAS1}
to evaluate the ASE result for 3GPP Case~2. Second, in order to show
the versatility of the studied 3GPP Case~1 with the \emph{linear}
LoS probability function shown in (\ref{eq:LoS_Prob_func_linear}),
as in~\cite{our_work_TWC2016}, we adopt the technique of approximating
the LoS probability function of 3GPP Case~2 shown in (\ref{eq:LoS_Prob_func_reverseS_shape})
by a 3-piece \emph{linear} function as
\begin{equation}
\textrm{Pr}^{\textrm{L}}\left(w\right)=\begin{cases}
\begin{array}{l}
1,\\
1-\frac{w-d_{1}}{d_{2}-d_{1}},\\
0,
\end{array} & \begin{array}{l}
L<w\leq d_{1}\\
d_{1}<w\leq d_{2}\\
w>d_{2}
\end{array}\end{cases},\label{eq:LoS_Prob_func_approx_reverseS}
\end{equation}
where $d_{1}$ and $d_{2}$ are set to 18.4\ m and 117.1\ m, respectively.
Note that $d_{1}$ is chosen as 18.4\ m because $\textrm{Pr}^{\textrm{L}}\left(d_{1}\right)=0.999\approx1$
in (\ref{eq:LoS_Prob_func_reverseS_shape}). Besides, the value of
$d_{2}$ is obtained from the requirement that $\textrm{Pr}^{\textrm{L}}\left(r\right)$
in (\ref{eq:LoS_Prob_func_approx_reverseS}) should go through the
point $\left(\frac{R_{1}}{\ln10},0.5\right)$, which is the crucial
point connecting the two segments in $\textrm{Pr}^{\textrm{L}}\left(r\right)$
of 3GPP Case~2 given by (\ref{eq:LoS_Prob_func_reverseS_shape}).
Note that the approximation of (\ref{eq:LoS_Prob_func_approx_reverseS})
can be easily improved by fitting the LoS probability function with
more than three pieces in (\ref{eq:LoS_Prob_func_approx_reverseS}).
For clarity, the combined case with the path loss function of (\ref{eq:PL_BS2UE_2slopes})
and the 3-piece LoS probability function of (\ref{eq:LoS_Prob_func_approx_reverseS})
is referred to as the Approximated 3GPP Case~2. Based on Theorem~\ref{thm:p_cov_UAS1},
we can readily extend the results in Appendix~E to analyze the Approximated
3GPP Case~2 in a tractable manner. The details are very similar to
those in~\cite{our_work_TWC2016} and thus omitted here for brevity.

\begin{figure}
\noindent \centering{}\includegraphics[width=8cm]{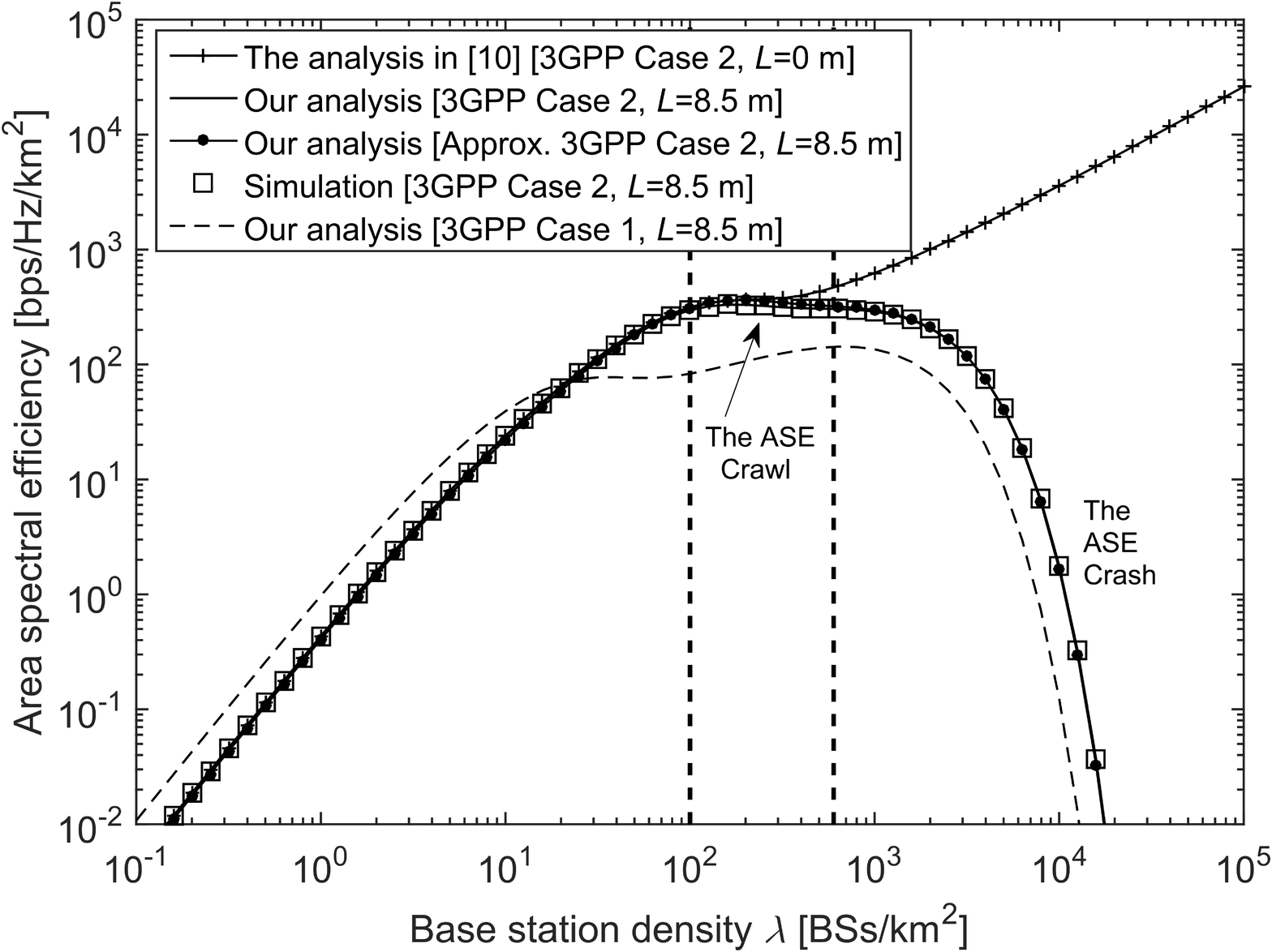}\renewcommand{\figurename}{Fig.}\caption{\label{fig:ASE_36828_3Dmod_fixedPower24dBm_gamma0dB_wBaseline}$A^{{\rm {ASE}}}\left(\lambda,\gamma_{0}\right)$
vs. $\lambda$ with $\gamma_{0}=0\,\textrm{dB}$ for 3GPP Case~2
(Rayleigh fading only).}
\end{figure}

In Fig.~\ref{fig:ASE_36828_3Dmod_fixedPower24dBm_gamma0dB_wBaseline},
we show the results of $A^{\textrm{ASE}}\left(\lambda,\gamma_{0}\right)$
for 3GPP Case~2.
 As can be seen from Fig.~\ref{fig:ASE_36828_3Dmod_fixedPower24dBm_gamma0dB_wBaseline},
the results of the Approximated 3GPP Case~2 based on (\ref{eq:LoS_Prob_func_approx_reverseS})
match those of 3GPP Case~2 well, thus showing the extensibility of
our analysis with the \emph{linear} LoS probability function. More
importantly, all the observations in Subsection~\ref{subsec:Sim-ASE-3GPP-Case-1}
are qualitatively valid for Fig.~\ref{fig:ASE_36828_3Dmod_fixedPower24dBm_gamma0dB_wBaseline}
except for some quantitative deviation. In more detail,
\begin{itemize}
\item The BS density range of \emph{the ASE Crawl} for 3GPP Case~2 is around
$\lambda\in\left[200,600\right]\,\textrm{BSs/km}^{2}$. And the ASE
also suffers from severe performance loss in UDNs, e.g., $\lambda>10^{3}\,\textrm{BSs/km}^{2}$.%
\item Compared with the baseline ASE Crash performance, the alternative
path loss model 3GPP Case~2 can delay \emph{the ASE Crash} from around
$\lambda=10^{4}\thinspace\textrm{BSs/km}^{2}$ to a bit more than
$\lambda=10^{4}\thinspace\textrm{BSs/km}^{2}$ when the ASE crashes
to 1\,$\textrm{bps/Hz/km}^{2}$.
\end{itemize}

\subsection{Key Takeaways\label{subsec:key-takeaways}}

The key takeaways of our study are summarized as follows,
\begin{itemize}
\item As explained in Subsection~\ref{subsec:The-ASE-Crash}, the fundamental
reason of \emph{the ASE crash} is rooted in the geometry of BS deployments.
Consequently, as shown in previous subsections, both the path loss
models and the multi-path fading models do not change the conclusion
of \emph{the ASE crash}. It should be noted that even if $L=0$, we
may still end up with $\bar{\gamma}\rightarrow1$ in (\ref{eq:SIR_toy_example}),
which leads to \emph{the ASE Crash}. One example is that if the BSs
are deployed on a circle or on the surface of a cylinder, then we
would have two base stations that are equally distant from the UE
located at the center. Another example is that in a very densely populated
pedestrian street where BSs cannot be placed anywhere on the street
itself, but only on walls abutting the street, we would also observe
two base stations having an equal distance from the UE standing in
the middle of the street~\cite{Gruber2016ASEcrash}. Compared with
the above two interesting examples, our paper analyzes a more likely
to happen scenario with random BS deployment, and establishes the
significant problem that these networks will face if the antenna height
issue is not considered in practice.
\item Regarding the solutions to avoid \emph{the ASE crash}, a straightforward
one is to lower the BS antenna height in 5G UDNs, so that the ASE
behavior of such networks would roll back to our previous results
in~\cite{our_GC_paper_2015_HPPP,our_work_TWC2016}. Such proposed
new BS deployment will allow to realize the potential gains of UDNs,
but needs a revolution on BS architectures and network deployment
in the future. Some new research challenges are as follows:%

\begin{itemize}
\item New measurement campaigns for the UE-height channels.
\item Futuristic BS architectures/hardware that are anti-vandalism/anti-theft/anti-hacking
at low-height positions.
\item New research avenues due to the emergence of fast time-variant shadow
fading due to random movement of UE-height objects, e.g., cars. Interesting
topics include new UE association strategies, agile power control,
fast link adaptation, etc.
\item Terrain-dependent network performance analysis considering hills,
elevated roads, etc.
\item New inter-BS communication means based on ground waves.
\end{itemize}
\item Another solution to mitigate the ASE crash is interference coordination.%
{} In order to mitigate inter-cell interference, LTE Release~8 incorporates
the inter-cell interference coordination (ICIC) features~\cite{Book_CoMP},
which provide mechanisms to coordinate RB usage among neighboring
BSs, e.g. high interference indicator (HII) and overload indicator
(OI) for the uplink and relative narrow-band transmit power (RNTP)
for the downlink. These features may be used in current macrocell
base stations, but not widely, due to overhead and delay issues in
the backhaul interface as well as the complexity of finding a good
solution using local measures. In practice, such features have not
been used in small cells, the topic of the paper. Moreover, it is
worth noting that LTE Release~13 incorporates a new small cell feature
to enable dynamic muting of small cell BSs~\cite{Book_CoMP} by the
use of discovery reference signals. Using such mechanism to mitigate
the overwhelming interference causing \emph{the ASE Crash} is a topic
for further study.
\item Another solution to mitigate the ASE crash is beam steering/shaping
using multi-antenna technologies. However, it should be noted that
sub-6GHz small cell products are of small form factor and targeted
at a low price. Thus, the number of antennas that can be placed in
the small cell BS is quite limited, which in turn limits such beam
steering/shaping capabilities. Such techniques are more beneficial
for millimeter wave solutions with a larger number of antennas, but
they are out of the scope of this paper, because this technology requires
a completely different modeling, as the 3GPP has done in the standardization
of the 5G networks~\cite{TR38.802}.%
{} In more detail, we need to consider new millimeter wave communication
features, such as short-range coverage, the blockage effect, very
low inter-cell interference, molecule absorption and re-radiation,
high Doppler shift, etc. Moreover, note that bringing beam steering
and beam shaping into the paper would mandate the study on the channel
correlation among different UEs. As a results, millimeter wave transmissions
with multiple antennas to mitigate \emph{the ASE Crash} are for further
study.
\end{itemize}

\section{Conclusion\label{sec:Conclusion}}

We presented a new and significant theoretical discovery, i.e., the
serious problem of \emph{the ASE Crash}. If the absolute height difference
between BS antenna and UE antenna is larger than zero, then the ASE
performance will continuously decrease toward \emph{zero} with the
network densification for UDNs. One way to overcome \emph{the ASE
Crash} is to lower the BS antenna height to the UE antenna height,
which will revolutionize the approach of BS architecture and network
deployment in the future. Other ways to counter-measure \emph{the
ASE Crash} could be pro-active muting of BSs, dynamic beam tracking,
cooperation of neighbouring BSs, and so on, which are worth further
studying in the context of UDNs.

\section*{Appendix~A: Proof of Theorem~\ref{thm:p_cov_UAS1}\label{sec:Appendix-A}}

In (\ref{eq:Theorem_1_p_cov}), $T_{n}^{{\rm {L}}}$ and $T_{n}^{{\rm {NL}}}$
are the components of the coverage probability for the case when the
signal comes from \emph{the $n$-th piece LoS path} and for the case
when the signal comes from \emph{the $n$-th piece NLoS path}, respectively.
The calculation of $T_{n}^{{\rm {L}}}$ is based on (\ref{eq:geom_dis_PDF_UAS1_LoS_thm})
and Lemma~\ref{lem:Pcov_LoS_and_NLoS_Rayleigh}. In (\ref{eq:geom_dis_PDF_UAS1_LoS_thm}),
$f_{R,n}^{{\rm {L}}}\left(r\right)$ characterizes the geometrical
density function of the typical UE with \emph{no other LoS BS} and
\emph{no NLoS BS} providing a better link to the typical UE than its
serving BS (a BS with \emph{the $n$-th piece LoS path}). The calculation
of $T_{n}^{{\rm {NL}}}$ is based on (\ref{eq:geom_dis_PDF_UAS1_NLoS_thm})
and Lemma~\ref{lem:Pcov_LoS_and_NLoS_Rayleigh}. The interpretation
of (\ref{eq:geom_dis_PDF_UAS1_NLoS_thm}) is similar to that for the
calculation of $T_{n}^{{\rm {L}}}$.

\section*{Appendix~B: Proof of Lemma~\ref{lem:Pcov_LoS_and_NLoS_Rayleigh}\label{sec:Appendix-B}}

In Lemma~\ref{lem:Pcov_LoS_and_NLoS_Rayleigh}, ${\rm {Pr}}\left[\frac{P\zeta_{n}^{{\rm {L}}}\left(\sqrt{r^{2}+L^{2}}\right)h}{I_{{\rm {agg}}}+P_{{\rm {N}}}}>\gamma\right]$
can be calculated as

${\rm {Pr}}\left[\frac{P\zeta_{n}^{{\rm {L}}}\left(\sqrt{r^{2}+L^{2}}\right)h}{I_{{\rm {agg}}}+P_{{\rm {N}}}}>\gamma\right]$
\begin{eqnarray}
\hspace{-0.3cm} & = & \hspace{-0.3cm}\mathbb{E}_{\left[I_{{\rm {agg}}}\right]}\left\{ \textrm{Pr}\left[h>\frac{\gamma\left(I_{{\rm {agg}}}+P_{{\rm {N}}}\right)}{P\zeta_{n}^{{\rm {L}}}\left(\sqrt{r^{2}+L^{2}}\right)}\right]\right\} \nonumber \\
\hspace{-0.3cm} & = & \hspace{-0.3cm}\mathbb{E}_{\left[I_{{\rm {agg}}}\right]}\left\{ \bar{F}_{H}\left(\frac{\gamma\left(I_{{\rm {agg}}}+P_{{\rm {N}}}\right)}{P\zeta_{n}^{{\rm {L}}}\left(\sqrt{r^{2}+L^{2}}\right)}\right)\right\} ,\label{eq:Pr_SINR_req_UAS1_LoS}
\end{eqnarray}
where $\mathbb{E}_{\left[X\right]}\left\{ \cdot\right\} $ denotes
the expectation operation taking the expectation over the variable
$X$ and $\bar{F}_{H}\left(h\right)$ denotes the CCDF of RV $h$.
Since we assume $h$ to be an exponential RV, we have $\bar{F}_{H}\left(h\right)=\exp\left(-h\right)$
and thus (\ref{eq:Pr_SINR_req_UAS1_LoS}) can be further derived as\vspace{0.2cm}

\noindent ${\rm {Pr}}\hspace{-0.1cm}\left[\frac{P\zeta_{n}^{{\rm {L}}}\left(\sqrt{r^{2}\hspace{-0.1cm}+\hspace{-0.1cm}L^{2}}\right)h}{I_{{\rm {agg}}}+P_{{\rm {N}}}}>\gamma\right]$

\noindent
\begin{eqnarray}
\hspace{-0.3cm} & = & \hspace{-0.3cm}\mathbb{E}_{\left[I_{{\rm {agg}}}\right]}\left\{ \bar{F}_{H}\left(\frac{\gamma\left(I_{{\rm {agg}}}+P_{{\rm {N}}}\right)}{P\zeta_{n}^{{\rm {L}}}\left(\sqrt{r^{2}+L^{2}}\right)}\right)\right\} \nonumber \\
\hspace{-0.3cm} & = & \hspace{-0.3cm}\exp\hspace{-0.1cm}\left(\hspace{-0.1cm}-\frac{\gamma P_{{\rm {N}}}}{P\zeta_{n}^{{\rm {L}}}\left(\sqrt{r^{2}+L^{2}}\right)}\hspace{-0.1cm}\right)\nonumber \\
\hspace{-0.3cm} &  & \hspace{-0.3cm}\times\mathbb{E}_{\left[I_{{\rm {agg}}}\right]}\hspace{-0.1cm}\left\{ \hspace{-0.1cm}\exp\hspace{-0.1cm}\left(\hspace{-0.1cm}-\frac{\gamma I_{{\rm {agg}}}}{P\zeta_{n}^{{\rm {L}}}\left(\sqrt{r^{2}+L^{2}}\right)}\hspace{-0.1cm}\right)\hspace{-0.1cm}\right\} \nonumber \\
\hspace{-0.3cm} & = & \hspace{-0.3cm}\exp\hspace{-0.1cm}\left(\hspace{-0.1cm}-\frac{\gamma P_{{\rm {N}}}}{P\zeta_{n}^{{\rm {L}}}\left(\sqrt{r^{2}+L^{2}}\right)}\hspace{-0.1cm}\right)\nonumber \\
\hspace{-0.3cm} & = & \hspace{-0.3cm}\times\mathscr{L}_{I_{{\rm {agg}}}}^{{\rm {L}}}\hspace{-0.1cm}\left(\hspace{-0.1cm}\frac{\gamma}{P\zeta_{n}^{{\rm {L}}}\left(\sqrt{r^{2}+L^{2}}\right)}\hspace{-0.1cm}\right)\hspace{-0.1cm},\hspace{-0.1cm}\label{eq:Pr_SINR_req_wLT_UAS1_LoS}
\end{eqnarray}

\noindent where $\mathscr{L}_{I_{{\rm {agg}}}}^{{\rm {L}}}\left(s\right)$
is the Laplace transform of RV $I_{{\rm {agg}}}$ evaluated at $s$
on the condition of the event that the typical UE is associated with
a BS with a LoS path. Note that $\exp\left(-\frac{\gamma P_{{\rm {N}}}}{P\zeta_{n}^{{\rm {L}}}\left(\sqrt{r^{2}+L^{2}}\right)}\right)$
measures the probability that \emph{the signal power exceeds the noise
power} by a factor of at least $\gamma$, and $\mathscr{L}_{I_{{\rm {agg}}}}^{{\rm {L}}}\left(\frac{\gamma}{P\zeta_{n}^{{\rm {L}}}\left(\sqrt{r^{2}+L^{2}}\right)}\right)$
measures the probability that \emph{the signal power exceeds the aggregate
interference power} by a factor of at least $\gamma$. Since $h$
follows an exponential distribution, the product of the above probabilities
yields the probability that \emph{the signal power exceeds the sum
power of the noise and the aggregate interference} by a factor of
at least $\gamma$. Based on the presented UAS, we can derive $\mathscr{L}_{I_{{\rm {agg}}}}^{{\rm {L}}}\left(s\right)$
as\vspace{0.2cm}

\noindent $\mathscr{L}_{I_{{\rm {agg}}}}^{{\rm {L}}}\left(s\right)$

\noindent
\begin{eqnarray}
\hspace{-0.3cm}\hspace{-0.3cm} & = & \hspace{-0.3cm}\mathbb{E}_{\left[I_{{\rm {agg}}}\right]}\left\{ \left.\exp\left(-sI_{{\rm {agg}}}\right)\right|B^{{\rm {L}}}\right\} \nonumber \\
\hspace{-0.3cm}\hspace{-0.3cm} & = & \hspace{-0.3cm}\mathbb{E}_{\left[\Phi,\left\{ \beta_{i}\right\} ,\left\{ g_{i}\right\} \right]}\left\{ \left.\exp\left(-s\sum_{i\in\Phi/b_{o}}P\beta_{i}\left(w\right)g_{i}\right)\right|B^{{\rm {L}}}\right\} .\label{eq:laplace_term_LoS_UAS1_general_seg_part1}
\end{eqnarray}
According to~\cite{Jeff2011}, $\mathbb{E}_{\left[g\right]}\left\{ \left.\exp\left(-sP\beta\left(w\right)g\right)\right|B^{{\rm {L}}}\right\} $
in (\ref{eq:laplace_term_LoS_UAS1_general_seg_part1}) should consider
interference from both LoS and NLoS paths. Thus, $\mathscr{L}_{I_{{\rm {agg}}}}^{{\rm {L}}}\left(s\right)$
can be further derived as

\noindent $\mathscr{L}_{I_{{\rm {agg}}}}^{{\rm {L}}}\left(s\right)$
\begin{eqnarray}
\hspace{-0.3cm}\hspace{-0.3cm} & = & \hspace{-0.3cm}\exp\left(-2\pi\lambda\int_{r}^{+\infty}\textrm{{\rm {Pr}}}^{{\rm {L}}}\left(\sqrt{u^{2}+L^{2}}\right)\right.\nonumber \\
\hspace{-0.3cm}\hspace{-0.3cm} &  & \hspace{-0.3cm}\times\left[1-\mathbb{E}_{\left[g\right]}\left\{ \exp\left(-sP\zeta^{{\rm {L}}}\left(\sqrt{u^{2}+L^{2}}\right)g\right)\right\} \right]udu\nonumber \\
\hspace{-0.3cm}\hspace{-0.3cm} &  & \hspace{-0.3cm}-2\pi\lambda\int_{r_{1}}^{+\infty}\left[1-\textrm{{\rm {Pr}}}^{{\rm {L}}}\left(\sqrt{u^{2}+L^{2}}\right)\right]\nonumber \\
\hspace{-0.3cm}\hspace{-0.3cm} &  & \hspace{-0.3cm}\left.\times\left[1-\mathbb{E}_{\left[g\right]}\left\{ \exp\left(-sP\zeta^{{\rm {NL}}}\left(\sqrt{u^{2}+L^{2}}\right)g\right)\right\} \right]udu\right)\nonumber \\
\hspace{-0.3cm}\hspace{-0.3cm} & = & \hspace{-0.3cm}\exp\left(-2\pi\lambda\int_{r}^{+\infty}\frac{\textrm{{\rm {Pr}}}^{{\rm {L}}}\left(\sqrt{u^{2}+L^{2}}\right)u}{1+\left(sP\zeta^{{\rm {L}}}\left(\sqrt{u^{2}+L^{2}}\right)\right)^{-1}}du\right)\nonumber \\
\hspace{-0.3cm}\hspace{-0.3cm} &  & \hspace{-0.3cm}\times\exp\left(-2\pi\lambda\int_{r_{1}}^{+\infty}\frac{\left[1-\textrm{{\rm {Pr}}}^{{\rm {L}}}\left(\sqrt{u^{2}+L^{2}}\right)\right]u}{1+\left(sP\zeta^{{\rm {NL}}}\left(\sqrt{u^{2}+L^{2}}\right)\right)^{-1}}du\right).\hspace{-0.3cm}\hspace{-0.3cm}\hspace{-0.3cm}\nonumber \\
\hspace{-0.3cm}\hspace{-0.3cm}\label{eq:laplace_term_LoS_UAS1_general_seg_part2}
\end{eqnarray}

Plugging $s=\frac{\gamma}{P\zeta_{n}^{{\rm {L}}}\left(\sqrt{r^{2}+L^{2}}\right)}$
into (\ref{eq:laplace_term_LoS_UAS1_general_seg_part2}) and further
plugging (\ref{eq:laplace_term_LoS_UAS1_general_seg_part2}) into
(\ref{eq:Pr_SINR_req_wLT_UAS1_LoS}), we can obtain the general expression
of ${\rm {Pr}}\left[\frac{P\zeta_{n}^{{\rm {L}}}\left(\sqrt{r^{2}+L^{2}}\right)h}{I_{{\rm {agg}}}+P_{{\rm {N}}}}>\gamma\right]$
shown in (\ref{eq:Pr_SINR_req_UAS1_LoS_thm}).

In a similar way, we can obtain the general expression of ${\rm {Pr}}\left[\frac{P\zeta_{n}^{{\rm {NL}}}\left(\sqrt{r^{2}+L^{2}}\right)h}{I_{{\rm {agg}}}+P_{{\rm {N}}}}>\gamma\right]$
shown in (\ref{eq:Pr_SINR_req_UAS1_NLoS_thm}), which concludes our
proof.%

\section*{Appendix~C: Proof of Theorem~\ref{thm:the_ASE_Falls_Theorem}\label{sec:Appendix-C}}

By applying the theory of limits on $p^{{\rm {cov}}}\left(\lambda,\gamma\right)$
derived in Theorem~\ref{thm:p_cov_UAS1}, we can obtain that $\underset{\lambda\rightarrow+\infty}{\lim}p^{{\rm {cov}}}\left(\lambda,\gamma\right)=\underset{\lambda\rightarrow+\infty}{\lim}T_{1}^{{\rm {L}}}+\underset{\lambda\rightarrow+\infty}{\lim}T_{1}^{{\rm {NL}}}$.
This is because
\begin{itemize}
\item When $\lambda\rightarrow+\infty$, the typical 2D distance $r$ from
the UE to its serving BS $b_{o}$ approaches zero, i.e., $r\rightarrow0$,
which should be dominantly characterized by either \emph{the first-piece
LoS path loss function} or \emph{the first-piece NLoS path loss function}
due to the short-distance link and $L$ is smaller than $d_{1}$ in
practical SCNs~\cite{SCM_pathloss_model,TR36.828}; and
\item According to Appendix~A, $T_{1}^{{\rm {L}}}$ and $T_{1}^{{\rm {NL}}}$
measure nothing but the components of the coverage probability for
the cases that the signal comes from \emph{the first-piece LoS path}
and that the signal comes from \emph{the first-piece NLoS path}, respectively.
\end{itemize}

Moreover, when $\lambda\rightarrow+\infty$, we have $\underset{\lambda\rightarrow+\infty}{\lim}T_{1}^{{\rm {NL}}}=0$
due to $\underset{\lambda\rightarrow+\infty}{\lim}f_{R,1}^{{\rm {NL}}}\left(r\right)=0$.
In more detail, $f_{R,1}^{{\rm {NL}}}\left(r\right)$ approaches zero
when $\lambda\rightarrow+\infty$, i.e., $r\rightarrow0$, because
\begin{itemize}
\item According to (\ref{eq:def_r_2}), $\underset{r\rightarrow0}{\lim}r_{2}=\underset{r_{2}}{\arg}\left\{ \zeta^{{\rm {L}}}\left(\sqrt{r_{2}^{2}+L^{2}}\right)=\zeta_{1}^{{\rm {NL}}}\left(L\right)\right\} \stackrel{\bigtriangleup}{=}r_{2}^{\textrm{min}}$,
which is a non-zero value due to the non-zero value of $L$.
\item Thus, the term $\exp\left(-\int_{0}^{r_{2}}{\rm {Pr}}^{{\rm {L}}}\left(\sqrt{u^{2}+L^{2}}\right)2\pi u\lambda du\right)$
in (\ref{eq:geom_dis_PDF_UAS1_NLoS_thm}) can be upper-bounded by
$\exp\left(-{\rm {Pr}}^{{\rm {L}}}\left(\sqrt{\left(r_{2}^{\textrm{min}}\right)^{2}+L^{2}}\right)\pi\lambda\left(r_{2}^{\textrm{min}}\right)^{2}\right)$,
which approaches zero when $\lambda\rightarrow+\infty$. Note that
we assume ${\rm {Pr}}^{{\rm {L}}}\left(w\right)$ to be a monotonically
decreasing function with respect to $w$ as explained in Section~\ref{sec:System-Model}.
\item Hence, we can obtain $\underset{\lambda\rightarrow+\infty}{\lim}f_{R,1}^{{\rm {NL}}}\left(r\right)=0$
considering the definition of $f_{R,1}^{{\rm {NL}}}\left(r\right)$
in (\ref{eq:geom_dis_PDF_UAS1_NLoS_thm}).
\end{itemize}

Therefore, we can claim that $\underset{\lambda\rightarrow+\infty}{\lim}p^{{\rm {cov}}}\left(\lambda,\gamma\right)=\underset{\lambda\rightarrow+\infty}{\lim}T_{1}^{{\rm {L}}}$,
which is in line with the intuitive fact that when $\lambda\rightarrow+\infty$,
the coverage probability should be mainly contributed by the case
that the signal comes from \emph{the first-piece LoS path}.

Next, we show that $\underset{\lambda\rightarrow+\infty}{\lim}{\rm {Pr}}\left[\frac{P\zeta_{1}^{{\rm {L}}}\left(\sqrt{r^{2}+L^{2}}\right)h}{I_{{\rm {agg}}}+N_{0}}>\gamma\right]=0$,
so that we can get $\underset{\lambda\rightarrow+\infty}{\lim}T_{1}^{{\rm {L}}}=0$.
From (\ref{eq:Pr_SINR_req_UAS1_LoS_thm}) and (\ref{eq:laplace_term_LoS_UAS1_general_seg_thm}),
we have\vspace{0.2cm}

\noindent ${\rm {Pr}}\left[\frac{P\zeta_{1}^{{\rm {L}}}\left(\sqrt{r^{2}+L^{2}}\right)h}{I_{{\rm {agg}}}+N_{0}}>\gamma\right]$
\begin{eqnarray}
\hspace{-0.3cm}\hspace{-0.3cm} & <\hspace{-0.3cm} & \exp\left(-2\pi\lambda\int_{r}^{+\infty}\frac{{\rm {Pr}}^{{\rm {L}}}\left(\sqrt{u^{2}+L^{2}}\right)u}{1+\left(\frac{\gamma P\zeta^{{\rm {L}}}\left(\sqrt{u^{2}+L^{2}}\right)}{P\zeta_{1}^{{\rm {L}}}\left(\sqrt{r^{2}+L^{2}}\right)}\right)^{-1}}du\right)\label{eq:p_cov_ineq_step1}\\
\hspace{-0.3cm}\hspace{-0.3cm} & <\hspace{-0.3cm} & \exp\left(-2\pi\lambda\int_{r}^{\tau r}\frac{{\rm {Pr}}^{{\rm {L}}}\left(\sqrt{u^{2}+L^{2}}\right)u}{1+\frac{1}{\gamma}\left(\frac{\sqrt{r^{2}+L^{2}}}{\sqrt{u^{2}+L^{2}}}\right)^{-\alpha_{1}^{{\rm {L}}}}}du\right)\label{eq:p_cov_ineq_step2}\\
\hspace{-0.3cm}\hspace{-0.3cm} & <\hspace{-0.3cm} & \exp\left(-2\pi\lambda\frac{{\rm {Pr}}^{{\rm {L}}}\left(\sqrt{\tau^{2}r^{2}+L^{2}}\right)\int_{r}^{\tau r}udu}{1+\frac{1}{\gamma}\left(\frac{\sqrt{r^{2}+L^{2}}}{\sqrt{\tau^{2}r^{2}+L^{2}}}\right)^{-\alpha_{1}^{{\rm {L}}}}}\right)\label{eq:p_cov_ineq_step3}\\
\hspace{-0.3cm}\hspace{-0.3cm} & \overset{r\rightarrow0}{<}\hspace{-0.3cm} & \exp\left(-\frac{{\rm {Pr}}^{{\rm {L}}}\left(L\right)\left(\tau^{2}-1\right)}{1+\frac{1}{\gamma}}\right)\overset{\Delta}{=}\kappa,\label{eq:p_cov_ineq}
\end{eqnarray}

\noindent where $\kappa$ is derived because
\begin{itemize}
\item As explained in Appendix~A, (\ref{eq:p_cov_ineq_step1}) only considers
the probability that the signal power beats the aggregate interference
power from all LoS BSs by a factor of at least $\gamma$, which over-estimates
the coverage probability. Besides, $s=\frac{\gamma}{P\zeta_{1}^{{\rm {L}}}\left(\sqrt{r^{2}+L^{2}}\right)}$
is plugged into (\ref{eq:p_cov_ineq_step1}).
\item In (\ref{eq:p_cov_ineq_step2}), we further concentrate on the LoS
interference that is relatively close to the UE, i.e., $u\in\left(r,\tau r\right],\left(1<\tau<+\infty\right)$.
Besides, \emph{the first-piece LoS path loss function} has also been
used for such LoS interference, because when $r\rightarrow0$ we have
$\tau r\rightarrow0$, which falls into the region that is dominantly
characterized by \emph{the first-piece LoS path loss function. }
\item We arrive at (\ref{eq:p_cov_ineq_step3}) because (i) $\left(\frac{\sqrt{r^{2}+L^{2}}}{\sqrt{u^{2}+L^{2}}}\right)^{-\alpha_{1}^{{\rm {L}}}}<\left(\frac{\sqrt{r^{2}+L^{2}}}{\sqrt{\tau^{2}r^{2}+L^{2}}}\right)^{-\alpha_{1}^{{\rm {L}}}},u\in\left(r,\tau r\right]$;
and (ii) ${\rm {Pr}}^{{\rm {L}}}\left(\sqrt{u^{2}+L^{2}}\right)>{\rm {Pr}}^{{\rm {L}}}\left(\sqrt{\tau^{2}r^{2}+L^{2}}\right),u\in\left(r,\tau r\right]$,
since ${\rm {Pr}}^{{\rm {L}}}\left(w\right)$ is a monotonically decreasing
function with respect to $w$ as explained in Section~\ref{sec:System-Model}.
\item Finally, we plug $\underset{\lambda\rightarrow+\infty}{\lim}{\rm {Pr}}^{{\rm {L}}}\left(\sqrt{\tau^{2}r^{2}+L^{2}}\right)={\rm {Pr}}^{{\rm {L}}}\left(L\right)$,
$\underset{\lambda\rightarrow+\infty}{\lim}\frac{\sqrt{r^{2}+L^{2}}}{\sqrt{\tau^{2}r^{2}+L^{2}}}=1$,
$\int_{r}^{\tau r}udu=\frac{1}{2}\left(\tau^{2}-1\right)r^{2}$ and
$\underset{\lambda\rightarrow+\infty}{\lim}\pi r^{2}\lambda=1$ into
(\ref{eq:p_cov_ineq_step3}) to obtain (\ref{eq:p_cov_ineq}). Note
that we have $\underset{\lambda\rightarrow+\infty}{\lim}\pi r^{2}\lambda=1$
because the typical coverage radius, denoted by $\hat{r}$, permits
$\underset{\lambda\rightarrow+\infty}{\lim}\pi\hat{r}^{2}\lambda=1$,
i.e., the typical coverage area $\pi\hat{r}^{2}$ is in the order
of $\frac{1}{\lambda}$. Hence, $\underset{\lambda\rightarrow+\infty}{\lim}\pi r^{2}\lambda=1$
holds because $r\rightarrow\hat{r}$ when $\lambda\rightarrow+\infty$.
\end{itemize}

Since $\tau$ takes an arbitrary and finite value, we have $\underset{\lambda\rightarrow+\infty}{\lim}{\rm {Pr}}\left[\frac{P\zeta_{1}^{{\rm {L}}}\left(\sqrt{r^{2}+L^{2}}\right)h}{I_{{\rm {agg}}}+N_{0}}>\gamma\right]=0$
because $\kappa$ can be reduced to any arbitrarily small value when
$\tau$ is sufficiently large. For example, when ${\rm {Pr}}^{{\rm {L}}}\left(L\right)=0.8$,
$\gamma=1$, and $\tau$ equals to a moderate value of 8, then $\kappa$
is smaller than $1\times10^{-11}$. Therefore, according to the definition
of $T_{1}^{{\rm {L}}}$ in (\ref{eq:Theorem_1_p_cov}), we can get
$\underset{\lambda\rightarrow+\infty}{\lim}T_{1}^{{\rm {L}}}=\underset{\lambda\rightarrow+\infty}{\lim}p^{{\rm {cov}}}\left(\lambda,\gamma\right)=0$.

Furthermore, since $\gamma$ is arbitrary for $p^{{\rm {cov}}}\left(\lambda,\gamma\right)$,
we can set $\gamma>\gamma_{0}$ and put the UE into a complete outage
in UDNs, i.e., $\underset{\lambda\rightarrow+\infty}{\lim}\textrm{Pr}\left[\mathrm{SINR}>\gamma_{0}\right]\equiv0$.
Since the UDNs are now operating below the minimum working SINR $\gamma_{0}$,
according to the definition of ASE in (\ref{eq:ASE_def}), we have
$\underset{\lambda\rightarrow+\infty}{\lim}A^{\textrm{ASE}}\left(\lambda,\gamma_{0}\right)=0$,
which completes our proof.

\section*{Appendix~D: Proof of Lemma~\ref{lem:Pcov_LoS_and_NLoS_Rician}\label{sec:Appendix-D}}

Let $SINR^{\textrm{L}}=\frac{P\zeta_{n}^{{\rm {L}}}\left(\sqrt{r^{2}+L^{2}}\right)h}{I_{{\rm {agg}}}+P_{{\rm {N}}}}$,
then ${\rm {Pr}}\left[\frac{P\zeta_{n}^{{\rm {L}}}\left(\sqrt{r^{2}+L^{2}}\right)h}{I_{{\rm {agg}}}+P_{{\rm {N}}}}>\gamma\right]$
can be reformulated as\vspace{0.2cm}

${\rm {Pr}}\left[SINR^{\textrm{L}}>\gamma\right]$
\begin{eqnarray}
\hspace{-0.3cm} & = & \hspace{-0.3cm}{\rm {Pr}}\left[\frac{1}{SINR^{\textrm{L}}}<\frac{1}{\gamma}\right]\nonumber \\
\hspace{-0.3cm} & = & \hspace{-0.3cm}\int_{0}^{\frac{1}{\gamma}}f_{\frac{1}{SINR^{\textrm{L}}}}\left(x\right)dx\nonumber \\
\hspace{-0.3cm} & \overset{(a)}{=} & \hspace{-0.3cm}\int_{0}^{\frac{1}{\gamma}}\frac{1}{2\pi}\mathcal{\int_{\mathrm{-}\infty}^{\infty}}\mathcal{F}_{\frac{1}{SINR^{\textrm{L}}}}(\omega)\exp\left(-\textrm{j}\omega x\right)d\omega dx\nonumber \\
\hspace{-0.3cm} & \overset{(b)}{=} & \hspace{-0.3cm}\int_{-\infty}^{+\infty}\left(\frac{1-\exp\left(\frac{\textrm{j}\omega}{\gamma}\right)}{2\pi\textrm{j}\omega}\right)\mathcal{F}_{\frac{1}{SINR^{\textrm{L}}}}(\omega)d\omega,\label{eq:Pr_SINR_req_UAS1_LoS_Rician}
\end{eqnarray}
where in ($a$) $\mathcal{F}_{\frac{1}{SINR^{\textrm{L}}}}(\omega)$
denotes the characteristic function of $\frac{1}{SINR^{\textrm{L}}}$
and in ($b$) the integral with respect to $x$ has been computed
to simplify the expression. Furthermore, $\mathcal{F}_{\frac{1}{SINR^{\textrm{L}}}}(\omega)$
can be written as%
\vspace{0.2cm}

\noindent $\mathcal{F}_{\frac{1}{SINR^{\textrm{L}}}}(\omega)$
\begin{eqnarray}
\hspace{-0.3cm}\hspace{-0.3cm} & \overset{(a)}{=} & \hspace{-0.3cm}\mathbb{E}_{\left[\frac{1}{SINR^{\textrm{L}}}\right]}\left\{ \exp\left(\textrm{j}\omega\frac{1}{SINR^{\textrm{L}}}\right)\right\} \nonumber \\
\hspace{-0.3cm}\hspace{-0.3cm} & \overset{(b)}{=} & \hspace{-0.3cm}\mathbb{E}_{\left[S^{{\rm {L}}},I_{{\rm {agg}}}^{{\rm {L}}},I_{{\rm {agg}}}^{{\rm {NL}}}\right]}\left\{ \exp\left(\textrm{j}\omega\frac{I_{{\rm {agg}}}^{{\rm {L}}}+I_{{\rm {agg}}}^{{\rm {NL}}}+P_{{\rm {N}}}}{S^{{\rm {L}}}}\right)\right\} \nonumber \\
\hspace{-0.3cm}\hspace{-0.3cm} & = & \hspace{-0.3cm}\mathbb{E}_{\left[S^{{\rm {L}}}\right]}\left\{ \mathbb{E}_{\left[I_{{\rm {agg}}}^{{\rm {L}}}\right]}\left\{ \exp\left(\textrm{j}\omega\frac{I_{{\rm {agg}}}^{{\rm {L}}}}{S^{{\rm {L}}}}\right)\right\} \right.\nonumber \\
\hspace{-0.3cm}\hspace{-0.3cm} & = & \hspace{-0.3cm}\left.\times\mathbb{E}_{\left[I_{{\rm {agg}}}^{{\rm {NL}}}\right]}\left\{ \exp\left(\textrm{j}\omega\frac{I_{{\rm {agg}}}^{{\rm {NL}}}}{S^{{\rm {L}}}}\right)\right\} \exp\left(\textrm{j}\omega\frac{P_{{\rm {N}}}}{S^{{\rm {L}}}}\right)\right\} \nonumber \\
\hspace{-0.3cm}\hspace{-0.3cm} & \overset{(c)}{=} & \hspace{-0.3cm}\mathbb{E}_{\left[h\right]}\hspace{-0.1cm}\left\{ \exp\hspace{-0.1cm}\left(\textrm{j}\omega\frac{P_{{\rm {N}}}}{P\zeta_{n}^{{\rm {L}}}\hspace{-0.1cm}\left(\sqrt{r^{2}\hspace{-0.1cm}+\hspace{-0.1cm}L^{2}}\right)\hspace{-0.1cm}h}\right)\right.\nonumber \\
\hspace{-0.3cm}\hspace{-0.3cm} &  & \hspace{-0.3cm}\times\mathbb{E}_{\left[\Phi^{{\rm {L}}},\left\{ \beta_{i}^{{\rm {L}}}\right\} ,\left\{ g_{i}^{{\rm {L}}}\right\} \right]}\hspace{-0.1cm}\left\{ \hspace{-0.1cm}\exp\hspace{-0.1cm}\left(\textrm{j}\omega\frac{\sum_{i\in\Phi^{{\rm {L}}}}P\beta_{i}^{{\rm {L}}}g_{i}^{{\rm {L}}}}{P\zeta_{n}^{{\rm {L}}}\hspace{-0.1cm}\left(\sqrt{r^{2}\hspace{-0.1cm}+\hspace{-0.1cm}L^{2}}\right)\hspace{-0.1cm}h}\right)\hspace{-0.1cm}\right\} \nonumber \\
\hspace{-0.3cm}\hspace{-0.3cm} &  & \hspace{-0.3cm}\left.\times\mathbb{E}_{\left[\Phi^{{\rm {NL}}}\hspace{-0.1cm},\left\{ \beta_{i}^{{\rm {NL}}}\right\} ,\left\{ g_{i}^{{\rm {NL}}}\right\} \right]}\hspace{-0.1cm}\left\{ \hspace{-0.1cm}\exp\hspace{-0.1cm}\left(\hspace{-0.1cm}\textrm{j}\omega\frac{\sum_{i\in\Phi^{{\rm {NL}}}}\hspace{-0.1cm}P\beta_{i}^{{\rm {NL}}}g_{i}^{{\rm {NL}}}}{P\zeta_{n}^{{\rm {L}}}\hspace{-0.1cm}\left(\sqrt{r^{2}\hspace{-0.1cm}+\hspace{-0.1cm}L^{2}}\right)\hspace{-0.1cm}h}\hspace{-0.1cm}\right)\hspace{-0.1cm}\right\} \hspace{-0.1cm}\right\} \hspace{-0.1cm},\label{eq: char_func_UAS1_LoS_Rician}
\end{eqnarray}
where
\begin{itemize}
\item ($a$) comes from the definition of $\mathcal{F}_{\frac{1}{SINR^{\textrm{L}}}}(\omega)$~\cite{Book_Integrals},
\item ($b$) breaks down the expression of $SINR^{\textrm{L}}$, where $S^{\textrm{L}}=P\zeta_{n}^{{\rm {L}}}\left(\sqrt{r^{2}+L^{2}}\right)h$
and $h$ follows a non-central chi-squared distribution due to our
assumption of \emph{Rician} fading for LoS transmissions~\cite{Book_Integrals},
$I_{{\rm {agg}}}^{{\rm {L}}}$ denotes the aggregated interference
from LoS interfering BSs to the typical UE, and $I_{{\rm {agg}}}^{{\rm {NL}}}$
denotes the aggregated interference from NLoS interfering BSs to the
typical UE,
\item ($c$) further breaks down the expression of $I_{{\rm {agg}}}^{{\rm {L}}}$
and $I_{{\rm {agg}}}^{{\rm {NL}}}$.%
{} To take $I_{{\rm {agg}}}^{{\rm {L}}}$ as an example, we have $I_{{\rm {agg}}}^{{\rm {L}}}=\sum_{i\in\Phi^{{\rm {L}}}}P\beta_{i}^{{\rm {L}}}g_{i}^{{\rm {L}}}$,
where $\Phi^{{\rm {L}}}$, $\beta_{i}^{{\rm {L}}}$ and $g_{i}^{{\rm {L}}}$
denote the set of LoS interfering BSs, the path loss of the $i$-th
LoS interfering BS and the \emph{Rician} fading power of the $i$-th
LoS interfering BS.
\end{itemize}
\noindent %

We define $ISR_{{\rm {aggL}}}^{{\rm {L}}}=\mathbb{E}_{\left[\Phi^{{\rm {L}}},\left\{ \beta_{i}^{{\rm {L}}}\right\} ,\left\{ g_{i}^{{\rm {L}}}\right\} \right]}\left\{ \exp\left(\textrm{j}\omega\frac{\sum_{i\in\Phi^{{\rm {L}}}}P\beta_{i}^{{\rm {L}}}g_{i}^{{\rm {L}}}}{P\zeta_{n}^{{\rm {L}}}\left(\sqrt{r^{2}+L^{2}}\right)h}\right)\right\} $,
which is the LoS interference to LoS signal ratio and can be derived
as%
\vspace{0.2cm}

\noindent $ISR_{{\rm {aggL}}}^{{\rm {L}}}$
\begin{eqnarray}
\hspace{-0.3cm} & \overset{(a)}{=} & \hspace{-0.3cm}\exp\left(-2\pi\lambda\int_{r}^{+\infty}\textrm{{\rm {Pr}}}^{{\rm {L}}}\left(\sqrt{u^{2}+L^{2}}\right)\right.\nonumber \\
\hspace{-0.3cm} &  & \hspace{-0.3cm}\left.\times\left(\hspace{-0.1cm}1-\mathbb{E}_{\left[g^{{\rm {L}}}\right]}\hspace{-0.1cm}\left\{ \hspace{-0.1cm}\exp\hspace{-0.1cm}\left(\textrm{j}\omega\frac{\zeta^{{\rm {L}}}\left(\sqrt{u^{2}\hspace{-0.1cm}+\hspace{-0.1cm}L^{2}}\right)g^{{\rm {L}}}}{\zeta_{n}^{{\rm {L}}}\left(\sqrt{r^{2}\hspace{-0.1cm}+\hspace{-0.1cm}L^{2}}\right)h}\right)\hspace{-0.1cm}\right\} \hspace{-0.1cm}\right)\hspace{-0.1cm}udu\right)\hspace{-0.1cm},\label{eq:ISR_LoS_interf_to_LoS_signal}
\end{eqnarray}
where ($a$) is derived according to Campbell's theorem~\cite{book_Haenggi}
and the fact that any LoS interfering BS should stay away from the
typical UE by at least $r$ (see Theorem~\ref{thm:p_cov_UAS1}).
Since we assume that Rician fading is associated with LoS transmissions,
$\mathbb{E}_{\left[g^{{\rm {L}}}\right]}\left\{ \exp\left(\textrm{j}\omega\frac{\zeta^{{\rm {L}}}\left(\sqrt{u^{2}+L^{2}}\right)g^{{\rm {L}}}}{\zeta_{n}^{{\rm {L}}}\left(\sqrt{r^{2}+L^{2}}\right)h}\right)\right\} $
in (\ref{eq:ISR_LoS_interf_to_LoS_signal}) can be further written
as\vspace{0.2cm}

\noindent $\mathbb{E}_{\left[g^{{\rm {L}}}\right]}\left\{ \exp\left(\textrm{j}\omega\frac{\zeta^{{\rm {L}}}\left(\sqrt{u^{2}+L^{2}}\right)g^{{\rm {L}}}}{\zeta_{n}^{{\rm {L}}}\left(\sqrt{r^{2}+L^{2}}\right)h}\right)\right\} $
\begin{eqnarray}
\hspace{-0.3cm} & = & \hspace{-0.3cm}\int_{0}^{+\infty}\exp\left(\textrm{j}\omega\frac{\zeta^{{\rm {L}}}\left(\sqrt{u^{2}+L^{2}}\right)x}{\zeta_{n}^{{\rm {L}}}\left(\sqrt{r^{2}+L^{2}}\right)h}\right)f_{{\rm {fad}}}^{{\rm {L}}}\left(x\right)dx,\label{eq:expectation_gL_LoS_sig}
\end{eqnarray}
where $f_{{\rm {fad}}}^{{\rm {L}}}\left(x\right)$ denotes the PDF
of the multi-path fading power for LoS transmissions. For the assumed
Rician fading in Lemma~\ref{lem:Pcov_LoS_and_NLoS_Rician}, $f_{{\rm {fad}}}^{{\rm {L}}}\left(x\right)$
follows a non-central chi-squared distribution given by (\ref{eq:PDF_Rician_lem})~\cite{Book_Integrals}.%
{} Plugging (\ref{eq:expectation_gL_LoS_sig}) and (\ref{eq:PDF_Rician_lem})
into (\ref{eq:ISR_LoS_interf_to_LoS_signal}), yields (\ref{eq:ISR_LoS_interf_to_LoS_signal_lem}).%

In a similar way, we define $ISR_{{\rm {aggNL}}}^{{\rm {L}}}=\mathbb{E}_{\left[\Phi^{{\rm {NL}}},\left\{ \beta_{i}^{{\rm {NL}}}\right\} ,\left\{ g_{i}^{{\rm {NL}}}\right\} \right]}\left\{ \exp\left(\textrm{j}\omega\frac{\sum_{i\in\Phi^{{\rm {NL}}}}P\beta_{i}^{{\rm {NL}}}g_{i}^{{\rm {NL}}}}{P\zeta_{n}^{{\rm {L}}}\left(\sqrt{r^{2}+L^{2}}\right)h}\right)\right\} $,
which is the NLoS interference to LoS signal ratio and can be derived
as (\ref{eq:ISR_NLoS_interf_to_LoS_signal_lem}). Then, we can plug
(\ref{eq:ISR_LoS_interf_to_LoS_signal_lem}) and (\ref{eq:ISR_NLoS_interf_to_LoS_signal_lem})
into (\ref{eq: char_func_UAS1_LoS_Rician}), which concludes our proof
for the computation of ${\rm {Pr}}\left[\frac{P\zeta_{n}^{{\rm {L}}}\left(\sqrt{r^{2}+L^{2}}\right)h}{I_{{\rm {agg}}}+P_{{\rm {N}}}}>\gamma\right]$.%

The derivation of ${\rm {Pr}}\left[\frac{P\zeta_{n}^{{\rm {NL}}}\left(\sqrt{r^{2}+L^{2}}\right)h}{I_{{\rm {agg}}}+P_{{\rm {N}}}}>\gamma\right]$
is very similar to that of ${\rm {Pr}}\left[\frac{P\zeta_{n}^{{\rm {L}}}\left(\sqrt{r^{2}+L^{2}}\right)h}{I_{{\rm {agg}}}+P_{{\rm {N}}}}>\gamma\right]$,
which is omitted for brevity.%

\section*{Appendix~E: An Approximation Technique for Evaluating Theorem~\ref{thm:p_cov_UAS1}
for 3GPP Case~1\label{sec:Appendix-E}}

For 3GPP Case~1, according to Theorem~\ref{thm:p_cov_UAS1}, $p^{{\rm {cov}}}\left(\lambda,\gamma\right)$
can be computed by plugging (\ref{eq:PL_BS2UE_2slopes}) and (\ref{eq:LoS_Prob_func_linear})
into (\ref{eq:geom_dis_PDF_UAS1_LoS_thm})-(\ref{eq:laplace_term_NLoS_UAS1_general_seg_thm}).
In order to increase the tractability of the results, we propose the
following approximation for $w=\sqrt{r^{2}+L^{2}}$ in (\ref{eq:geom_dis_PDF_UAS1_LoS_thm})-(\ref{eq:laplace_term_NLoS_UAS1_general_seg_thm}),
\begin{equation}
w\approx\tilde{w}=\begin{cases}
\begin{array}{l}
L,\\
\frac{r+L}{\sqrt{2}},\\
r,
\end{array} & \hspace{-0.3cm}\begin{array}{l}
0\leq r\leq v_{1}\\
v_{1}\leq r\leq v_{2}\\
r>v_{2}
\end{array},\end{cases}\label{eq:approx_dis_BS2UE}
\end{equation}
where $v_{1}=\left(\sqrt{2}-1\right)L$ and $v_{2}=\left(\sqrt{2}+1\right)L$.
Such approximation is based on the following three lower bounds of
$w$:
\begin{itemize}
\item $w=\sqrt{r^{2}+L^{2}}\geq L$, which is tight when $r$ is very small,
i.e., $0\leq r\leq v_{1}$.
\item $w=\sqrt{r^{2}+L^{2}}\geq\frac{r+L}{\sqrt{2}}$, which is tight when
$r$ is relatively small, i.e., $v_{1}\leq r\leq v_{2}$.
\item $w=\sqrt{r^{2}+L^{2}}\geq r$, which is tight when $r$ is relatively
large, i.e., $r>v_{2}$.
\end{itemize}
\noindent The above three lower bounds meet at $v_{1}$ and $v_{2}$,
which are defined as the switch points in (\ref{eq:approx_dis_BS2UE}).
For example, when $L=8.5$\,m, it is easy to verify that the maximum
absolute error of the approximation is merely around 1.5\,m.%
{} In the following, we show the computation of $T_{1}^{\textrm{L}}$
as an example on how to obtain $\left\{ T_{n}^{Path}\right\} $ in
$p^{{\rm {cov}}}\left(\lambda,\gamma\right)$ using the proposed approximation
of $w\approx\tilde{w}$ in (\ref{eq:approx_dis_BS2UE}).

From Theorem~\ref{thm:p_cov_UAS1}, for the range of $0<r\leq\sqrt{d_{1}^{2}-L^{2}}$,
$T_{1}^{\textrm{L}}$ can be calculated by
\begin{eqnarray}
\hspace{-0.3cm}T_{1}^{\textrm{L}}\hspace{-0.3cm} & = & \hspace{-0.3cm}\int_{0}^{\sqrt{d_{1}^{2}-L^{2}}}\exp\left(-\frac{\gamma\left(\sqrt{r^{2}+L^{2}}\right)^{\alpha^{{\rm {L}}}}P_{{\rm {N}}}}{PA^{{\rm {L}}}}\right)\nonumber \\
\hspace{-0.3cm} &  & \hspace{-0.3cm}\times\mathscr{L}_{I_{{\rm {agg}}}}^{{\rm {L}}}\left(\frac{\gamma\left(\sqrt{r^{2}+L^{2}}\right)^{\alpha^{{\rm {L}}}}}{PA^{{\rm {L}}}}\right)f_{R,1}^{{\rm {L}}}\left(r\right)dr,\label{eq:T_1_UAS1_LoS}
\end{eqnarray}
where $\zeta_{1}^{\textrm{L}}\left(w\right)=A^{{\rm {L}}}w^{-\alpha^{{\rm {L}}}}$
from (\ref{eq:PL_BS2UE_2slopes}) has been plugged into (\ref{eq:T_1_UAS1_LoS})
and $\mathscr{L}_{I_{{\rm {agg}}}}^{{\rm {L}}}\left(s\right)$ is
the Laplace transform of $I_{{\rm {agg}}}$ for the LoS signal transmission
evaluated at $s$.

In (\ref{eq:T_1_UAS1_LoS}), according to Theorem~\ref{thm:p_cov_UAS1}
and (\ref{eq:LoS_Prob_func_linear}), $f_{R,1}^{\textrm{L}}\left(r\right)$
can be derived as

\noindent $f_{R,1}^{\textrm{L}}\left(r\right)$
\begin{eqnarray}
\hspace{-0.3cm}\hspace{-0.3cm} & = & \hspace{-0.3cm}\exp\left(-\int_{0}^{r_{1}}\lambda\frac{\sqrt{u^{2}+L^{2}}}{d_{1}}2\pi udu\right)\nonumber \\
\hspace{-0.3cm}\hspace{-0.3cm} &  & \hspace{-0.3cm}\times\exp\left(-\int_{0}^{r}\lambda\left(1-\frac{\sqrt{u^{2}+L^{2}}}{d_{1}}\right)2\pi udu\right)\nonumber \\
\hspace{-0.3cm}\hspace{-0.3cm} &  & \hspace{-0.3cm}\times\left(1-\frac{\sqrt{r^{2}+L^{2}}}{d_{1}}\right)2\pi r\lambda\nonumber \\
\hspace{-0.3cm}\hspace{-0.3cm} & = & \hspace{-0.3cm}\exp\left(-\frac{2\pi\lambda}{3d_{1}}\left(\left(r_{1}^{2}+L^{2}\right)^{\nicefrac{3}{2}}-L^{3}\right)\right)\nonumber \\
\hspace{-0.3cm}\hspace{-0.3cm} &  & \hspace{-0.3cm}\times\exp\left(-\pi\lambda r^{2}+\frac{2\pi\lambda}{3d_{1}}\left(\left(r^{2}+L^{2}\right)^{\nicefrac{3}{2}}-L^{3}\right)\right)\nonumber \\
\hspace{-0.3cm}\hspace{-0.3cm} &  & \hspace{-0.3cm}\times\left(1-\frac{\sqrt{r^{2}+L^{2}}}{d_{1}}\right)2\pi r\lambda,\left(0<r\leq\sqrt{d_{1}^{2}-L^{2}}\right),\label{eq:spec_geom_dis_PDF_UAS1_LoS_seg1}
\end{eqnarray}
where $r_{1}=\sqrt{\left(\frac{A^{{\rm {NL}}}}{A^{{\rm {L}}}}\right)^{\frac{2}{\alpha^{{\rm {NL}}}}}\left(r^{2}+L^{2}\right)^{\frac{\alpha^{{\rm {L}}}}{\alpha^{{\rm {NL}}}}}-L^{2}}$
according to~(\ref{eq:def_r_1}).

Besides, according to Theorem~\ref{thm:p_cov_UAS1}, $\mathscr{L}_{I_{{\rm {agg}}}}^{{\rm {L}}}\left(s\right)$
in (\ref{eq:T_1_UAS1_LoS}) for the range of $0<r\leq\sqrt{d_{1}^{2}-L^{2}}$
should be computed by plugging (\ref{eq:PL_BS2UE_2slopes}) and (\ref{eq:LoS_Prob_func_linear})
into (\ref{eq:laplace_term_LoS_UAS1_general_seg_thm}). Similar to~\cite{our_work_TWC2016},
we need to break the integration interval into several segments according
to (\ref{eq:approx_dis_BS2UE}) and repeatedly calculate the following
two definite integrals: $Q_{1}=\int_{a}^{b}\frac{u}{1+\left(sPA^{Path}\right)^{-1}\left(\sqrt{u^{2}+L^{2}}\right)^{\alpha^{Path}}}du,$
and $Q_{2}=\int_{a}^{b}\frac{\sqrt{u^{2}+L^{2}}u}{1+\left(sPA^{Path}\right)^{-1}\left(\sqrt{u^{2}+L^{2}}\right)^{\alpha^{Path}}}du,$
where the string variable $Path$ takes the value of ``L'' and ``NL''
for the LoS and the NLoS cases, respectively. %

\bibliographystyle{IEEEtran}
\bibliography{Ming_library}

\end{document}